\documentclass[aps,preprint,nofootinbib,superscriptaddress,groupedaddress]{revtex4}
\usepackage{graphicx}
\usepackage{epstopdf}

\usepackage{amsmath,amsthm,latexsym,amssymb,amsfonts,epsfig}
\usepackage{fontenc,dsfont,layout}
\usepackage{hyperref}
\usepackage{psfrag}
\usepackage[cp1250]{inputenc}

\usepackage[usenames,dvipsnames]{xcolor}

\newcommand{\be}{\begin{equation}}
\newcommand{\ee}{\end{equation}}
\newcommand{\bea}{\begin{eqnarray}}
\newcommand{\eea}{\end{eqnarray}}

\begin{document}
\title{\boldmath Inflation in Supergravity from Field Redefinitions}

\author{Micha{\l} Artymowski}
\email{Michal.Artymowski@fuw.edu.pl}
\affiliation{Physics Department, Ariel University, Ariel 40700, Israel}
\author{Ido Ben-Dayan}
\email{Ido.Bendayan@gmail.com}
\affiliation{Physics Department, Ariel University, Ariel 40700, Israel}

\begin{abstract}
Supergravity (SUGRA) theories are specified by a few functions, most
notably the real K\"ahler function denoted by $G(T_i, \bar {T}_i) = K + \log |W|^2$, where K is
a real K\"ahler potential, and W is a holomorphic superpotential. A field redefinition
$T_i \rightarrow f_1(T_i)$ changes neither the theory nor the K\"ahler geometry. Similarly, 
the K\"ahler transformation, $K \rightarrow K + f_2 + \bar f_2, W \rightarrow e^{-f_2} W$  where $f_2$ is holomorphic and leaves G and hence the theory and the geometry invariant. However, if we perform a field redefinition only in $K(T_i,\bar{T}_i) \rightarrow K(f(T_i),f(\bar{T}_i))$, while keeping the same superpotential $W(T_i)$, we get a different theory, as G is not invariant under such a transformation while maintaining the same K\"ahler geometry. This freedom of choosing $f(T_i)$ allows construction of an infinite number of new theories given a fixed K\"ahler geometry and a predetermined superpotential W. Our construction generalizes previous ones that were limited by the holomorphic property of $W$. In particular, it~allows for novel inflationary SUGRA models and particle phenomenology model building, where~the different models correspond to different choices of field redefinitions. We~demonstrate this possibility by constructing several prototypes of inflationary models (hilltop, Starobinsky-like, plateau, log-squared and bell-curve) all in flat K\"ahler geometry and an originally renormalizable superpotential $W$. The models are in accord with current observations and predict $r\in[10^{-6},0.06]$ spanning several decades that can be easily obtained. In the bell-curve model, there~also exists a built-in gravitational reheating mechanism with $T_R\sim \mathcal{O}( 10^7 GeV)$.
\end{abstract}

\maketitle

\section{Introduction}

Cosmic inflation~\cite{Starobinsky:1980te,Lyth:1998xn,Sato:1980yn} is a hypothetical period of accelerated expansion of the Universe. Inflation solves certain problems of classical cosmology~\cite{Guth:1980zm} and is responsible for generating the primordial inhomogeneities of the Universe. Predictions of inflation are also highly consistent with experimental data~\cite{Akrami:2018odb}. Inflation usually requires flat potentials or flat directions in field space (for single and multi-field inflationary models respectively), which appear naturally in Supersymmetry (SUSY) and Supergravity (SUGRA). On~the other hand, the~well-known eta-problem is most clearly evident in SUGRA, where canonical K\"ahler potential of the form $K=T\bar{T}$ immediately generates a very steep potential along the radial direction of the field due to the exponential nature of the scalar potential $V=e^K(\cdots)$. In~the last decade many SUGRA inflationary models have been developed, with~the emphasis on the $\alpha$-attractors~\cite{Kallosh:2013yoa,Artymowski:2016pjz,Dimopoulos:2016yep}, for~which a non-canonical K\"ahler potential generates a kinetic term with a pole and, in~consequence, a~scalar potential with an inflationary~plateau. 

$\alpha$-attractors are characterized by K\"ahler potentials with non-zero K\"ahler curvature.~Such~a non-canonical  K\"ahler potential generally assumes some knowledge of the UV theory, and~restricts the possible UV completions. An~alternative to this approach is to investigate K\"ahler potentials of flat geometry with $\mathcal{R}_{i \bar j k \bar l} = 0$. For~instance,~\cite{Kawasaki:2000yn}
\begin{equation}
K_{\pm} = \pm \frac{1}{2}\left( T \pm \bar{T} \right)^2 + S\bar{S} . \label{eq:KahlerSimple}
\end{equation}
 Such a K\"ahler potential gives the scalars canonical kinetic terms and can always be viewed as a low energy approximation of some UV complete model without restrictions. Obviously, such a K\"ahler potential has a flat K\"ahler direction along $T_R\equiv \Re{T}$ $(T_I\equiv \Im{T})$ for $-$ $(+)$ in \eqref{eq:KahlerSimple}. One can use this flat direction to generate inflation. The~crucial ingredient is the shift symmetry of the K\"ahler potential, $T\rightarrow T\pm c$ for pure imaginary (real) $c$. Generalization of this form of K\"ahler potentials has been used throughout the literature. Generalizations beyond the simple shift symmetry of singlets to Higgs doublets and other symmetry groups have been suggested in~\cite{BenDayan:2010yz,Nakayama:2010sk}. 

The existing SUGRA literature connects the possibility and the type of model of inflation with the K\"ahler geometry. Perhaps the well-known examples are necessary conditions on the sectional curvature of the K\"ahler manifold~\cite{Badziak:2008yg,BenDayan:2008dv,Covi:2008cn}. It is, therefore, tempting to consider inflationary models as different classes of K\"ahler geometries. Obviously, this is not the case, since for a fixed K\"ahler manifold and metric, one can have various superpotentials leading to different models of inflation. A~rather generic construction was specified in \cite{Kallosh:2010xz}, where the authors considered $K = K_-$ and its generalizations with a superpotential of the form $W=S\, F(T)$, which give rise to a stable inflation trajectory of the form $V=F(\phi/\sqrt{2})^2$ (where $\phi$ is a real part of the field $T$). The~parameters of the superpotential $W$ still need to fulfill the slow-roll conditions, and~$F(T)$ is a holomorphic function. We wish to generalize the above construction and remove the limitation of holomorphism of the function $F$. 
The suggestion is the following. In~SUGRA, the~theory is specified, by~a few functions. The~true gauge invariant one for our discussion is the real K\"ahler function ({We work with natural units} $M_{pl}^{-2} = 8\pi G = 1$):
\be
G(T_i, \bar {T}_i) = K(T_i, \bar {T}_i) + \log |W(T_i)|^2 \, ,
\ee
where $K(T_i, \bar {T}_i)$ is the real  K\"ahler potential and $W(T_i)$ is the holomorphic superpotential. $G$ fixes the K\"ahler geometry and the full action can be written in terms of $G$ and its derivatives without the artificial separation between $K$ and $W$. Using $G$ one defines a K\"ahler geometry, defined by $g_{i \bar{j}} = G_{i\bar{j}} = \frac{\partial G}{\partial T_i \partial T^\dagger_j}$. The~holomorphic sectional curvature defined by $R_{i\bar{j}k\bar{l}} = G_{i\bar{j}k\bar{l}}-g^{m\bar{n}} G_{i k\bar{n}} G_{\bar{j}\bar{l}m}$ plays a crucial role in determining the stability of the model, since the Hessian matrix $V_{IJ}$ (i.e., the~matrix of the second derivatives of the scalar potential) is positively define? for~\cite{Covi:2008cn}
\begin{equation}
R_{i\bar{j}k\bar{l}} f^i f^{\bar{j}} f^k f^{\bar{l}} < \frac{2}{3}\frac{1}{1+\gamma}\, ,
\end{equation}
where $f^i = G^i/\sqrt{G^j G_j}$ and $\gamma \simeq H^2/m_{3/2}^2$. Please note that the sectional curvature depends only on $K$. The~particular example of \eqref{eq:KahlerSimple} gives a flat K\"ahler geometry, since in such a case $R_{i\bar{j}k\bar{l}} = 0$. $G$ is invariant under a K\"ahler transformation $K \rightarrow K + f_2 + \bar f_2, W \rightarrow e^{-f_2} W$, where $f_2(T_i)$ is holomorphic, so such a transformation does not change the physics. Field redefinitions  $T_i \rightarrow f_i(T_i)$ also do not change the physics. 
However, for~this to hold, this field redefinition must be applied to the full action, otherwise a new theory is constructed. Field redefinitions are in general not limited by holomorphism, and~non-holomorphic functions such as logarithm or a square root may also be considered. 
Our~suggestion is that given a theory with a fixed $K,W$ ($\equiv$ fixed G), we perform a field redefinition $T_i \rightarrow f_i(T_i)$ to $K$ only. As~a result, we get a new theory that allows us to construct various new inflationary models in~SUGRA. 

Similar models in prior work required for instance, the~use of non-canonical K\"ahler/non-flat K\"ahler geometry, or~were limited by the holomorphicity of the superpotential. We show that by judiciously choosing $T_i \rightarrow f_i(T_i)$, we can get various models, all with a simple fixed superpotential $W$, and~a flat K\"ahler geometry. In~particular, we will show that for a single K\"ahler  $K_{\pm} = \pm \frac{1}{2}\left( f(T) \pm f(\bar{T}) \right)^2 + S\bar{S}$ and superpotential 
$W=\Lambda S \, T$ we can generate small field, large field and Starobinsky type inflation, the~sole difference being the field redefinition that we picked. \mbox{Hence, classes of} inflation models differ neither by their K\"ahler geometry, nor by their superpotential, but~only by the field redefinition of the K\"ahler potential that we have specified. Let us stress that this method is applicable for any $G$ and therefore for any SUGRA theory, any K\"ahler geometry and any superpotential $W$. However, to~demonstrate the method's effectiveness and power we limit ourselves to the basic model of flat K\"ahler and a simple renormalizable superpotential. Analyzed~inflationary scenarios consist of both small and large field models, with~$10^{-6} \lesssim r \lesssim 0.06 $ and $n_s$ consistent with observational data. In~the bell-curve model, there also exists a  built-in gravitational reheating mechanism with $T_R\sim \mathcal{O}( 10^7 GeV)$.

One could ask how to test these ideas could be tested beyond the cosmological paradigm. One~can connect this research to the contemporary theory of topological materials. In~certain modern condensed matter systems there is important role of effective gravitational fields.  In~principle, this~opens the possibility to check in laboratory conditions certain ideas of quantum cosmology including the one presented here. For~more details on this issue see~\cite{Volovik:2014eca,Gorbar:2013dha,Parrikar:2014usa,Horava:2005jt}, and~the seminal book "The Universe in a Helium Droplet"~\cite{Volovik:2003fe}, by~Volovik.

The paper is organized as follows. In~Section~\ref{sec:general} we discuss the general method. In~Section~\ref{sec:prototypes} we give several prototypical examples of our method.  As~a consequence, we obtain theories with scalar potentials such as a monomial potential (Section~\ref{sec:monomial}), Mexican hat potential (Section~\ref{sec:local}), Starobinsky~inflation (Section~\ref{sec:starobinsky}), $\log^2\phi$ model (Section~\ref{sec:log}), plateau models (Section~\ref{sec:simple} and \ref{sec:modular}) and finally an exponential bell-curve potential (Section~\ref{sec:bell}) with an $\epsilon \sim N^{-n}$ parametrization of the slow-roll parameter. Finally, we~conclude in Section~\ref{sec:concl}. 

\section{Transformation of the K\"ahler~potential}\label{sec:general}

Consider the following field redefinition for \eqref{eq:KahlerSimple}:
\begin{equation}
K_{\pm} = \pm \frac{1}{2}\left( f\left(T\right) \pm f\left(\bar{T}\right) \right)^2 +S\bar{S} \label{eq:Kahler} \, ,
\end{equation}
where $f(T)$ has a dimension of mass. Let us stress that $T\rightarrow f(T)$ in the full action is a simple field redefinition and does not change the physics. The~novelty here is performing the transformation only on the K\"ahler potential. Hence, the~geometry is unchanged and in \eqref{eq:KahlerSimple} field space is flat. Our purpose here is to use it to construct new SUGRA models that support inflation. For~the superpotential of the form of
\begin{equation}
W = \Lambda \,  S \,  w(T) 
\end{equation}
one finds the F-term scalar potential from the equation
\begin{equation}
V = e^G \left( K^{T\bar{T}}G_T G_{\bar{T}} + K^{S\bar{S}}G_S G_{\bar{S}} -3 \right) \, ,
\end{equation}
where
\begin{equation}
G = K + \log|W|^2 \, ,\qquad G_T = \frac{\partial G}{\partial T} \, , \qquad G_S = \frac{\partial G}{\partial S} \, .
\end{equation}

The $S$ field is the so-called stabilizer and plays no other role.
The dependence on $S$ allows us to integrate it out supersymmetrically, giving a VEV $S = \bar{S} = 0$, yielding:
\begin{equation}
V =e^K|W_S|^2= \Lambda^2\exp\left(\pm\frac{1}{2} \left(f\left(T\right)\pm f\left(\bar{T}\right)\right)^2\right) |w(T)|^2  \, . \label{eq:V}
\end{equation}

To obtain a canonical K\"ahler metric let us introduce the following variable
\begin{equation}
U = f\left(T\right) \qquad \Rightarrow \qquad T = f^{-1}\left(U\right) \, ,
\end{equation}
which finally gives
\begin{eqnarray}
V_{+} &=& \Lambda^2 e^{2U_R^2}\left|w\left(f^{-1}\left(U\right)\right)\right|^2 \quad \text{for} \quad K = K_+ \, , \nonumber\\
V_{-} &=& \Lambda^2 e^{2U_I^2}\left|w\left(f^{-1}\left(U\right)\right)\right|^2 \quad \text{for} \quad K = K_- \, ,\label{eq:Vgeneral}
\end{eqnarray}
where $U= (U_R + i \, U_I) = (u_R + i\, u_I)/\sqrt{2}$ and $u_R$ and $u_I$ are real fields with canonical kinetic terms. Please note that after the field redefinition one finds $K_{U\bar{U}} = 1$ and the K\"ahler curvature remains zero. In~this work we investigate models with several forms of $f(T)$ and we show how they can be used to generate inflation. Inflation happens along the flat K\"ahler direction, which means that during inflation one finds
\begin{eqnarray}
V_{+} &=& \Lambda^2 \left|w\left(f^{-1}\left(i u_I/\sqrt{2}\right)\right)\right|^2, \nonumber\\
V_{-} &=& \Lambda^2 \left|w\left(f^{-1}\left(u_R/\sqrt{2}\right)\right)\right|^2 \, .\label{eq:Vpm}
\end{eqnarray}

In further part of this work we will assume $K = K_-$, unless~explicitly stated otherwise. This comes from the fact that for the considered range of models $K_-$ always generates inflation with sufficiently flat potential and graceful exit, which is not the case for $K = K_+$.  As~mentioned, we will focus on the simplest case of the superpotential $W = \Lambda S \, T$, for~which one finds
\begin{equation}
V = V_- = \Lambda^2 \left|f^{-1}\left(u_R/\sqrt{2}\right)\right|^2  . \label{eq:Vinfl}
\end{equation}

 As a consequence of the canonical kinetic terms for $u_R$ and $u_I$ the equation of motion of the inflaton field takes its well know form
\begin{equation}
\ddot{\phi} + 3H \dot{\phi} + V_\phi \, ,
\end{equation}
where $H$ is a Hubble parameter, $\phi = u_R$ or $u_I$ is the inflaton field and $V_\phi = \frac{dV}{d\phi}$. During~inflation one assumes that $\ddot{\phi} \ll H\dot{\phi}, V_\phi$, which is the {\emph slow-roll} approximation. In~such a case
\begin{equation}
\dot{\phi} \simeq -\frac{V_\phi}{3H} \, .
\end{equation}

Inflation ends when the absolute value of one of the following slow-roll parameters $\epsilon$ and $\eta$ becomes of order of 1, where
\begin{equation}
\epsilon = \frac{1}{2}\left(\frac{V_\phi}{V}\right)^2 \, , \qquad \eta = \frac{V_{\phi\phi}}{V} \, .
\end{equation}

Finally, in~this work we will compare the predictions of the considered inflationary models with the data~\cite{Akrami:2018odb}, which requires the obtaining of two quantities, which define main properties of the power spectrum of primordial inhomogeneities, namely tensor-to-scalar ratio $r$ and a spectral index $n_s$
\begin{equation}
r = 16\epsilon  \, ,\qquad n_s = 1 - 6\epsilon + 2\eta \, .
\end{equation}
$r$ and $n_s$ should be taken at a moment, when the pivot scale leaves the horizon.~This~moment corresponds to $N_\star$ e-folds before the end of inflation, where $N_\star$ is usually taken to be around $50$ or $60$. In~all the figures, the~green region corresponds to values of observables allowed by current~observations.

{Any SUGRA inflationary theory of the form of \eqref{eq:Vpm} can be also embedded in theories of modified gravity. The~duality between Einstein frame potentials and SUGRA inflationary models can be found in~\cite{Galante:2014ifa,Dalianis:2018afb}. Nevertheless, let us emphasize that the SUGRA models may have a unique thermal history of the Universe~\cite{Dalianis:2018afb}, which may lead to a different moment of horizon crossing of the pivot scale (i.e.,~to~a different $N_\star$). This can help to distinguish between inflationary models embedded in SUGRA or in theories of modified  {gravity} \cite{Dalianis:2018afb}.} 

\section{Prototypes of~Models} \label{sec:prototypes}
\vspace{-6pt}

\subsection{Monomial~Models} \label{sec:monomial}
Consider $U=f(T)=T^p$ for $p\geq 1$ and $w(T)=T$.
In such case the scalar potential for the canonically normalized field $U$ will be:
\be
V=\Lambda^2e^{u_I^2}\left(\frac{1}{2}\left(u_R^2+u_I^2\right)\right)^{1/p} ,
\ee
where inserting the correct powers of $M_{pl}$ to account for the correct dimensionality is obvious. It is also obvious that $u_I=0$ is an extremum and the global minimum is at $u_I=u_R=0$. Hence inflation takes place along the $u_I=0$ direction and the potential is simplified to a {well-known} monomial model:
\be
V=\Lambda^2 2^{-1/p}u_R^{2/p}  .
\ee

The predictions of such models are well known with
\be
n_s-1\simeq -\frac{1+p}{p N_\star},\quad r\simeq \frac{8}{p N_\star} \, .
\ee

As shown in~\cite{Akrami:2018odb} the predictions of the monomial model lies within the $2\sigma$ regime the Planck/BICEP data only for $N_\star = 50$ and $p\simeq 3$. This model is the only case considered in this paper, for~which $K_-$ and $K_+$ gives exactly the same inflationary potential. {This model is a first example of a theory, for~which the standard Kahler potential from~\cite{Kallosh:2010xz} could not generate considered potential. For~$K = -\frac{1}{2}(T-\bar{T})^2$ one would require $w \propto T^{1/p}$, which is explicitly non-holomorphic.}

\subsection{Locally Flat~Potentials} \label{sec:local}

The successful inflation requires only $\sim 60$ e-folds, one can safely assume that an inflationary potential may only be locally flat, just like in the case of hilltop models~\cite{Boubekeur:2005zm,Ben-Dayan:2013fva,BenDayan:2009kv}. In~this section, we~present the K\"ahler potential, for~which the local flatness appears around a local maximum of the potential, which is a stationary point of the order of $p-1$. Let us assume that
\begin{equation}
f(T) = \frac{M}{\sqrt{2}} \left( 1-\left(1-T\right)^{\frac{1}{p}}\right) \, , \qquad  w = T \, .
 \label{eq:modelLflat}
\end{equation}

After carrying out the procedure specified in the previous section, one finds the scalar potential of the form of
\begin{eqnarray}
V = \Lambda^2 e^{u_I^2} \left(1+\left(\frac{\sqrt{(M-u_R)^2+u_I^2}}{M}\right)^{2 p}+ \right.\nonumber\\
\left.-2 \left(\frac{\sqrt{(M-u_R)^2+u_I^2}}{M}\right)^p \cos \left(p \arctan\left(\frac{u_I}{M-u_R}\right)\right)\right) \, .
\end{eqnarray}

Inflation takes place along the flat K\"ahler direction, i.e.,~for $u_I = 0$, which gives
\begin{equation}
V(u_I = 0) =   \Lambda^2 \left(1 - \left(\frac{|M-u_R|}{M}\right)^p\right)^{2} \, . \label{eq:VeffLflat}
\end{equation}

\begin{figure}
\centering
\includegraphics[width=0.45\textwidth]{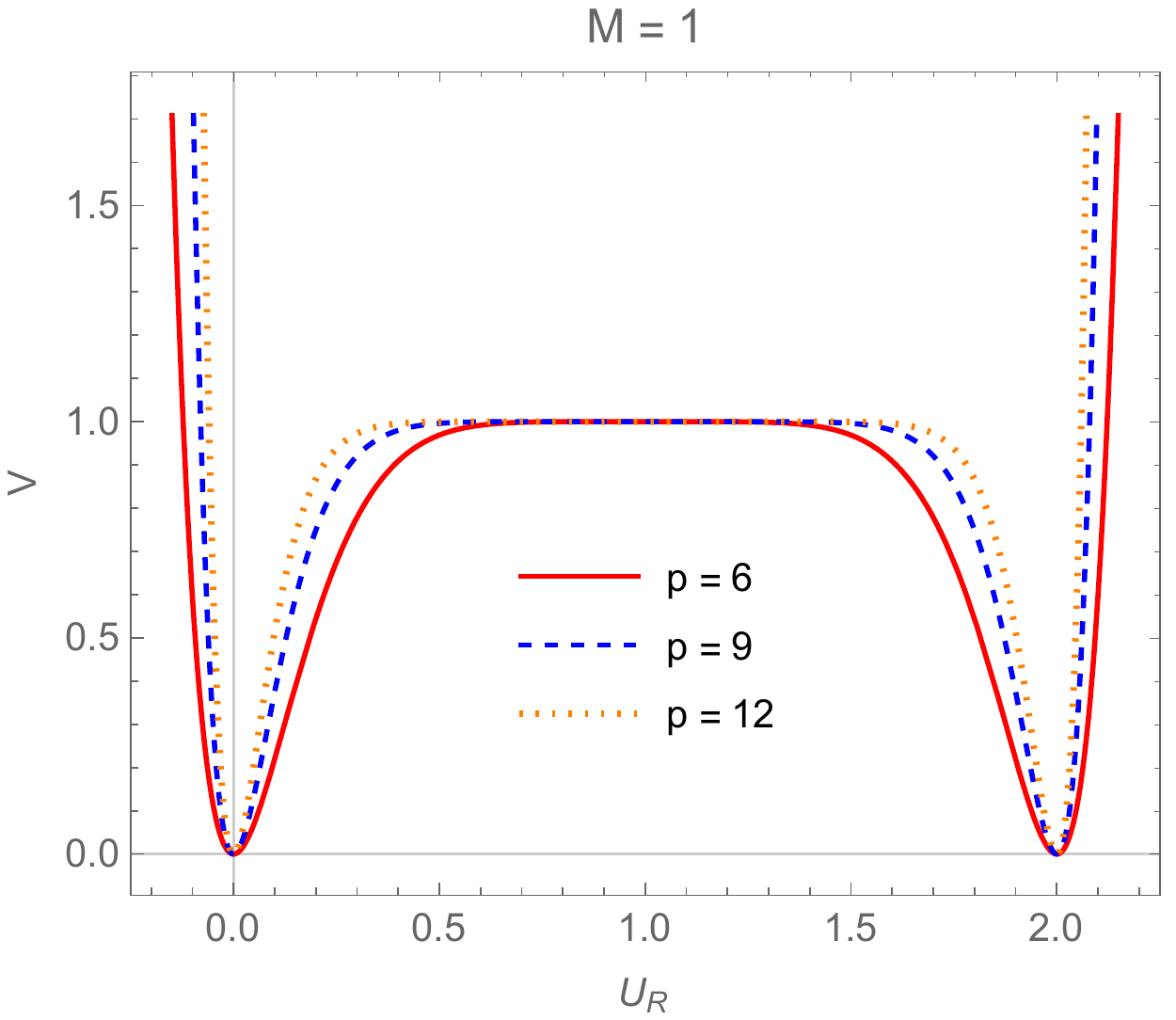}
\hspace{0.5cm}
\includegraphics[width=0.45\textwidth]{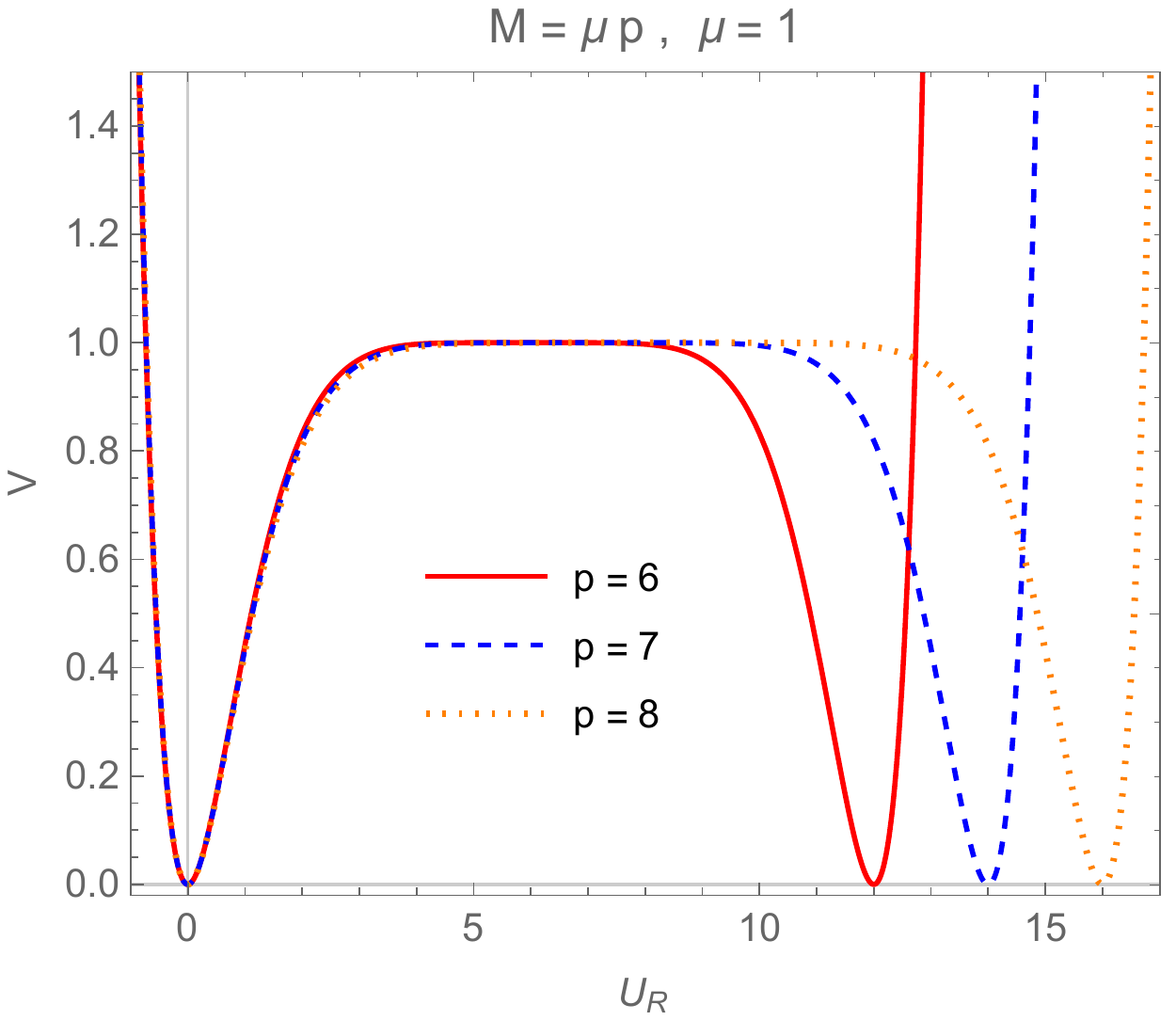}
\caption{ {Both} panels show the potential of the \eqref{eq:VeffLflat} model for different values of $p$ for $M = 1$ and $M = p \, \mu$ (left and right panels respectively), where $M$ and $\mu$ are $p$-independent constants. Inflation takes place around the local maximum of the potential. Right panel:  {Note} how the local maximum is moving towards bigger $u_R$ for bigger $p$.}
\label{fig:VLflat}
\end{figure}

The potential \eqref{eq:VeffLflat} is plotted in Figure~\ref{fig:VLflat}. Around the minimum at $u_R = 0$ one finds $V \simeq p^2 \Lambda^2 u_R^2/M^2$. We want to emphasize that the desired form of the inflationary potential may in this case only be obtained with $K = K_-$.

\begin{figure}
\centering
\includegraphics[width=0.45\textwidth]{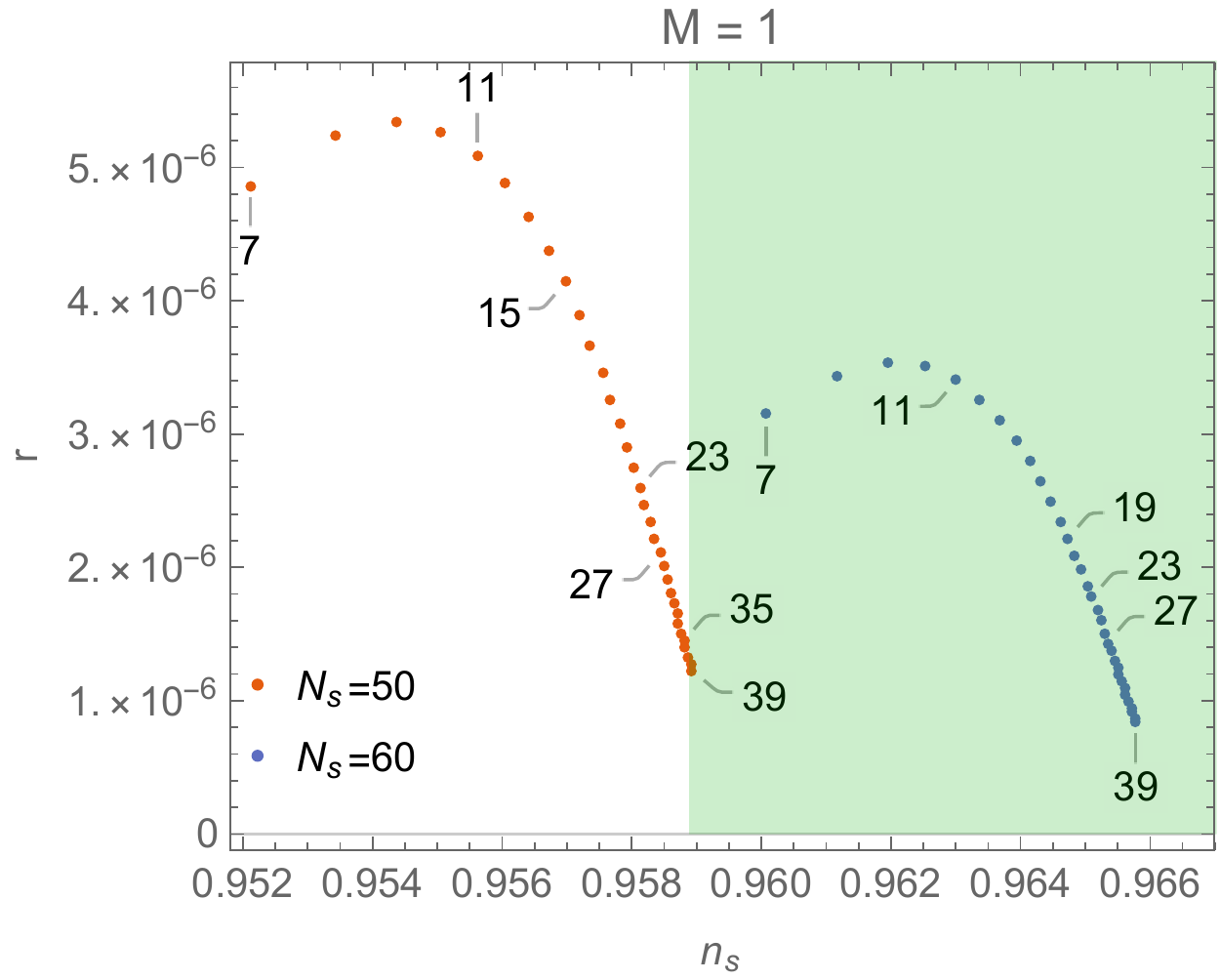}
\hspace{0.5cm}
\includegraphics[width=0.45\textwidth]{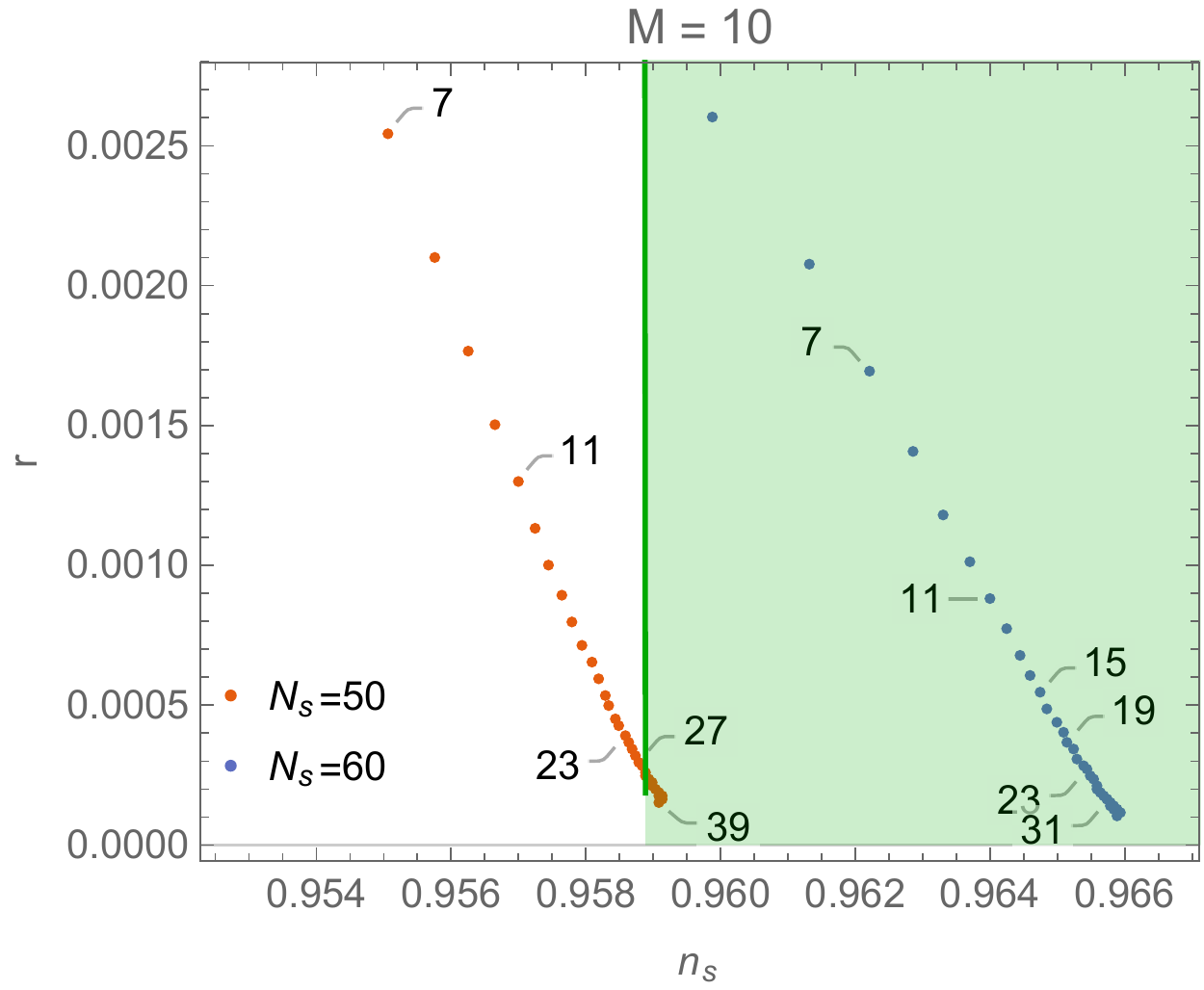}
\caption{ {Left} Panel: results for the model \eqref{eq:VeffLflat} model for $M = 1$, $p \in (7,39)$ and $N_\star = 50$ and $N_\star = 60$ (orange and blue dots respectively). Right panel: The same results for $M = 10$. Note how the predicted value of $r$ rise up with $M$.}
\label{fig:VLflatrns}
\end{figure}

The potential \eqref{eq:VeffLflat} has a local maximum at $u_R=0$, around which one finds a locally flat area of the potential. The~$u_R = 0$ is a stationary point of the theory, thus this model can be understood as a mixture between a hilltop inflation and a higher order saddle-point inflation. In~principle, as~shown in~\cite{Hamada:2015wea}, models with stationary point of the order of $p$ tend to give low scale of inflation together with $n_s \simeq 1 - \frac{2p}{(p-1)N_\star}$, which gives correct value of $n_s$ for $p \gtrsim 5$. The~results for the \eqref{eq:VeffLflat} model are plotted in the Figure~\ref{fig:VLflatrns}. Note how increasing $M$ increases $r$ and therefore the scale of inflation and field excursion, which can be estimate as
\begin{equation}
\Delta u_R \sim M \left(\frac{1}{2}-\frac{p}{70}\right) \, .
\end{equation}

Clearly, the~model $M = 1$ ($M = 10$) is a good example of a small (large) field~model. 

The \eqref{eq:modelLflat} model can be generalized into $f(T) = \frac{p}{\sqrt{2}}\mu (1-(1-T)^{1/p})$, which is equivalent to the~\eqref{eq:modelLflat} for $M = \mu \, p$. In~such a case the position of the local maximum is $p$-dependent.  {For more details about locally flat potentials and their relation to $\alpha$-attractors see}~\cite{Artymowski:2016pjz} {. For~more models on SUGRA hilltop inflation see}~\cite{Pinhero:2019sbg}. 

\unskip

\unskip

\subsection{Generalization of the Starobinsky~Inflation} \label{sec:starobinsky}

For
\begin{equation}
f(T) = - \frac{M}{\sqrt{2}} \log \left( 1-T\right) \, , \qquad  w = T \, ,\label{eq:Staro}
\end{equation}
one finds the scalar potential of the form of
\begin{equation}
 V = \Lambda^2 e^{u_I^2} \left(1-2 e^{-\frac{u_R}{M}} \cos \left(\frac{u_I}{M}\right)+e^{-2\frac{u_R}{M}}\right) \, . \label{eq:VStarobinsky}
\end{equation}

The potential obtains its global minimum for $u_I = u_R = 0$, while the inflation happens for $u_I = 0$ and $u_R > M$. In~such a case the potential \eqref{eq:VStarobinsky} takes the following approximate form
\begin{equation}
V = \Lambda^2  \left(1-e^{-\frac{u_R}{M}}\right)^{2} \, , \label{eq:VStarobinskyapp}
\end{equation}
which is a simple generalization of Starobinsky inflation. The~model has already been investigated in e.g.,~\cite{Kallosh:2014rga,Kallosh:2013xya} {and which can be defined by the  following Jordan frame action of the scalar-tensor { theory}
\begin{equation}
    S = \int d^4x \sqrt{-g}\left(\frac{\varphi}{2}R-\frac{\omega}{\varphi}(\partial\varphi)^2-\Lambda^2(\varphi-1)^2\right) \, ,
\end{equation}
where $\omega$ and $\Lambda$ are positive constants. Every model within the scalar-tensor theory may be also discussed in the Einstein frame, for~which the metric tensor is transformed into $g^E_{\mu\nu} = \varphi g_{\mu\nu}$. In~the Einstein frame the gravity is minimally coupled to the scalar field $\varphi$. After~the  field redefinition $\phi = \sqrt{\beta/2}\log(\varphi)$ one obtains the { action} 
\begin{equation}
    S = \int d^4x\sqrt{-g^E}\left(\frac{1}{2}R^E - \frac{1}{2}(\partial\phi)^2-\Lambda^2(1-e^{-\sqrt{2/\beta}\phi})\right) \, .
\end{equation}

Clearly, this model is fully consistent with \eqref{eq:VStarobinskyapp}.}

\begin{figure}
\centering
\includegraphics[width=0.45\textwidth]{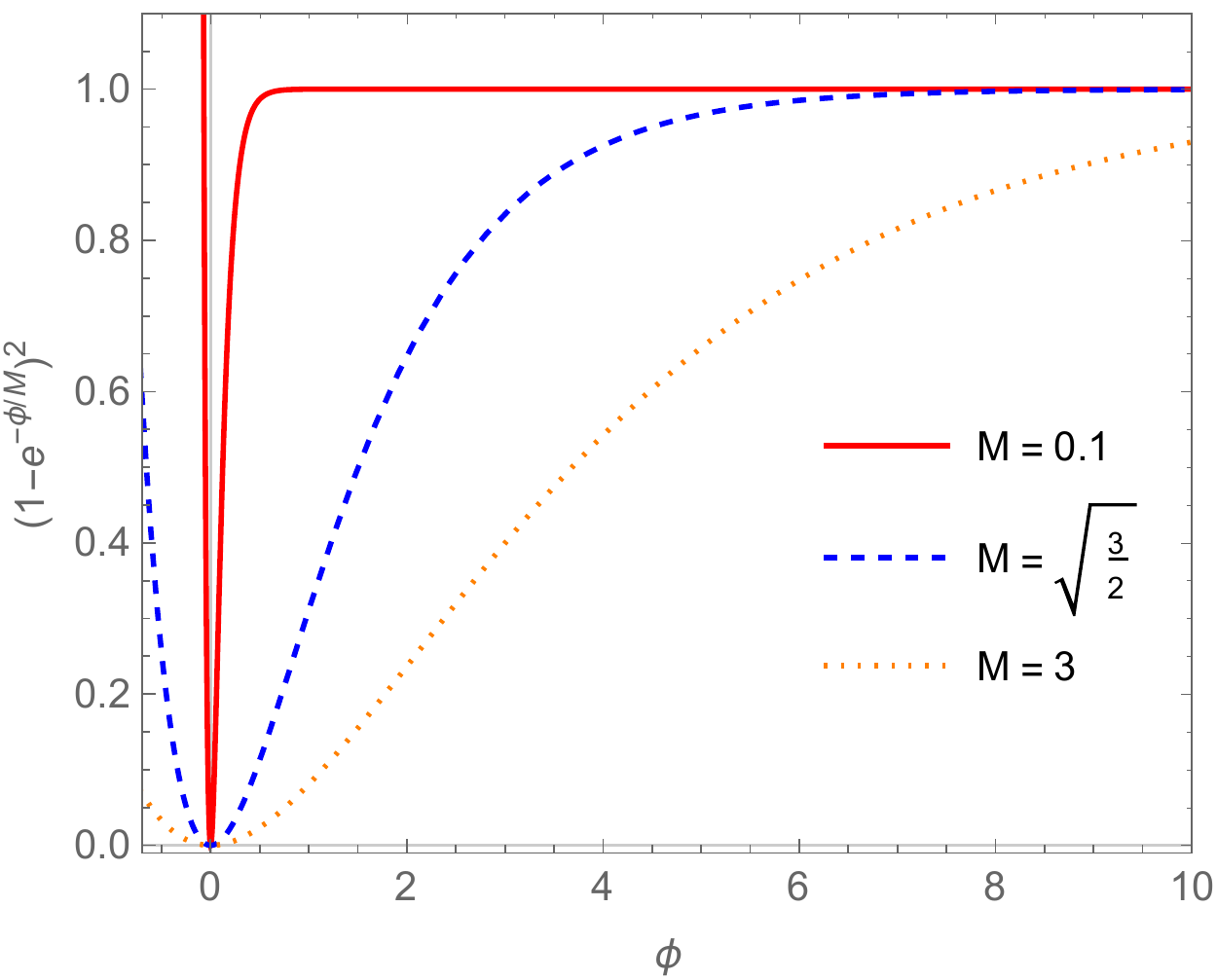}
\hspace{0.5cm}
\includegraphics[width=0.45\textwidth]{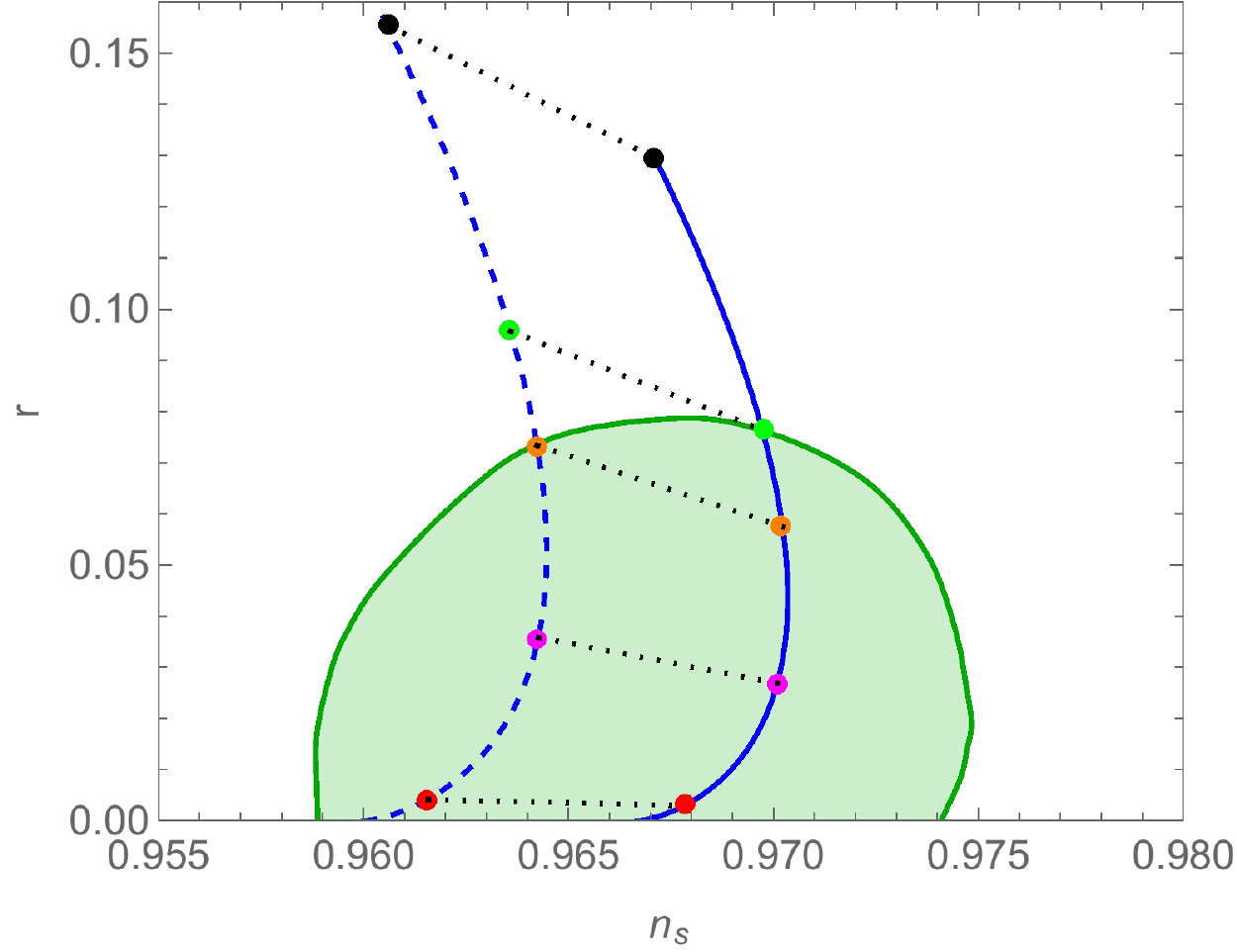}
\caption{Left panel:  Inflationary potential for different values of $M$. For~small $M$ the last $60$ e-folds of inflation happens on the plateau, while for $M \gg 1$ the inflationary potential resembles the $V\propto \phi^2$ model. Right panel: $r(n_s)$ for $N_\star = 50$ and $N_\star = 60$ (dashed and solid lines respectively). Red, pink, orange and green points corresponds to $M = \sqrt{3/2}$, $M = 5$, $M =10 $ and $M = 50$ respectively. The~black points correspond to $r = 4(1-n_s)$, which is the result of the $V \propto \phi^2$ model. For~$N_\star = 50$ and $N_\star = 60$ one obtains the consistency with the data for $M < 11$ and $M < 17.5$ respectively.}
\label{fig:Staro}
\end{figure}

In the $M \lesssim 1$ limit the results of this model correspond to $\alpha$-attractors, namely
\begin{equation}
r \simeq 8\frac{M^2}{N_\star^2} \, , \qquad n_s \simeq 1 - \frac{2}{N_\star} \, ,
\end{equation}
while in the $M \gg 1$ limit the model becomes equivalent to the $V \propto \phi^2$ inflation, which gives $r = 8/N_\star$ and $n_s = 1-2/N_\star$. The~inflationary potential for different values of $M$ together with the consistency of the model with the data have been presented in Figure~\ref{fig:Staro}. In~particular, for~$M = \sqrt{3/2}$ or $M = \sqrt{\beta/2}$ one recovers Starobinsky inflation or its Brans-Dicke generalization, respectively. $\beta$ in the Brans-Dicke theory is a parameter defined by the properties of kinetic term of the scalaron and {\emph not} by mass of fields,  just like in~\cite{Tsujikawa:2013ila}.

Please note that this model is a certain limit of the \eqref{eq:VeffLflat} scenario. If~in \eqref{eq:VeffLflat} one would consider $M = \mu\,  p$, then \eqref{eq:VeffLflat} would be equivalent to \eqref{eq:VStarobinskyapp} for $p \to \infty$. In~such a case the stationary point moves towards $u_R \to \infty$.

A similar model has been investigated in \cite{Pinhero:2018xbk}, where the authors consider $K = K_-$ and $f(T) = \arctan(T)$. In~such a case one recovers the scalar potential equivalent to $\alpha$-attractors, which give predictions very similar to the Starobinsky inflation. 
The \eqref{eq:Staro} model is a good example on how a choice of $K = K_+$ may give radically different results. In~such a case one finds
\begin{equation}
V = 2\Lambda^2\left(1-\cos\left(\frac{u_I}{M}\right) \right)\, ,
\end{equation}
which is a well-known potential of natural inflation~\cite{Adams:1992bn}, which, unlike the Starobinsky inflation, is inconsistent with the data~\cite{Akrami:2018odb}.
\unskip
\subsection{The Log$^2\phi$ Model for $K = K_-$} \label{sec:log}

Following the general model defined in the Section~\ref{sec:general} let us consider the following theory
\begin{equation}
f = \frac{M}{\sqrt{2}} e^{T} \, ,\qquad w =  T \, , \label{eq:modelLog2}
\end{equation}
which from \eqref{eq:Vgeneral} gives the following form of the scalar potential
\begin{equation}
V = \Lambda^2 e^{u_{I}^2} \left( \frac{1}{4}\log^2 \left(\frac{u_R^2 + u_I^2}{M^2}\right) + \arctan^2 \left( \frac{u_I}{u_R} \right) \right) \, . \label{eq:VS0}
\end{equation}

 {Inflation} takes place in the valley of $u_I = 0$. Global minima of the potential are $u_I = 0$ and $u_R = \pm M$, for~which one finds $V = 0$. The~potential along the inflationary trajectory reads
\begin{equation}
V (u_I = 0) = \Lambda^2 \log^{2} \left(\frac{|u_R|}{M}\right) \, . \label{eq:VLognKm}
\end{equation}

The entire evolution of the field may be described as reaching the $u_I = 0$ valley from random initial conditions and then following the valley until $u_R$ reach global minimum at $u_R = \pm M$. The~results for such a model line plotted in  {Figure}~\ref{fig:LogKm} (dashed line). {Again, we want to emphasize that the $\log^2(|u_R|/M)$ potential could not be obtained using the standard $K = -\frac{1}{2}(T-\bar{T})^2$ from~\cite{Kallosh:2010xz}, since $W \propto \log(T)$ is non-holomorphic.} 

\subsubsection{The $K = K_+$ Scenario}

\begin{figure}
\centering
\includegraphics[width=0.42\textwidth]{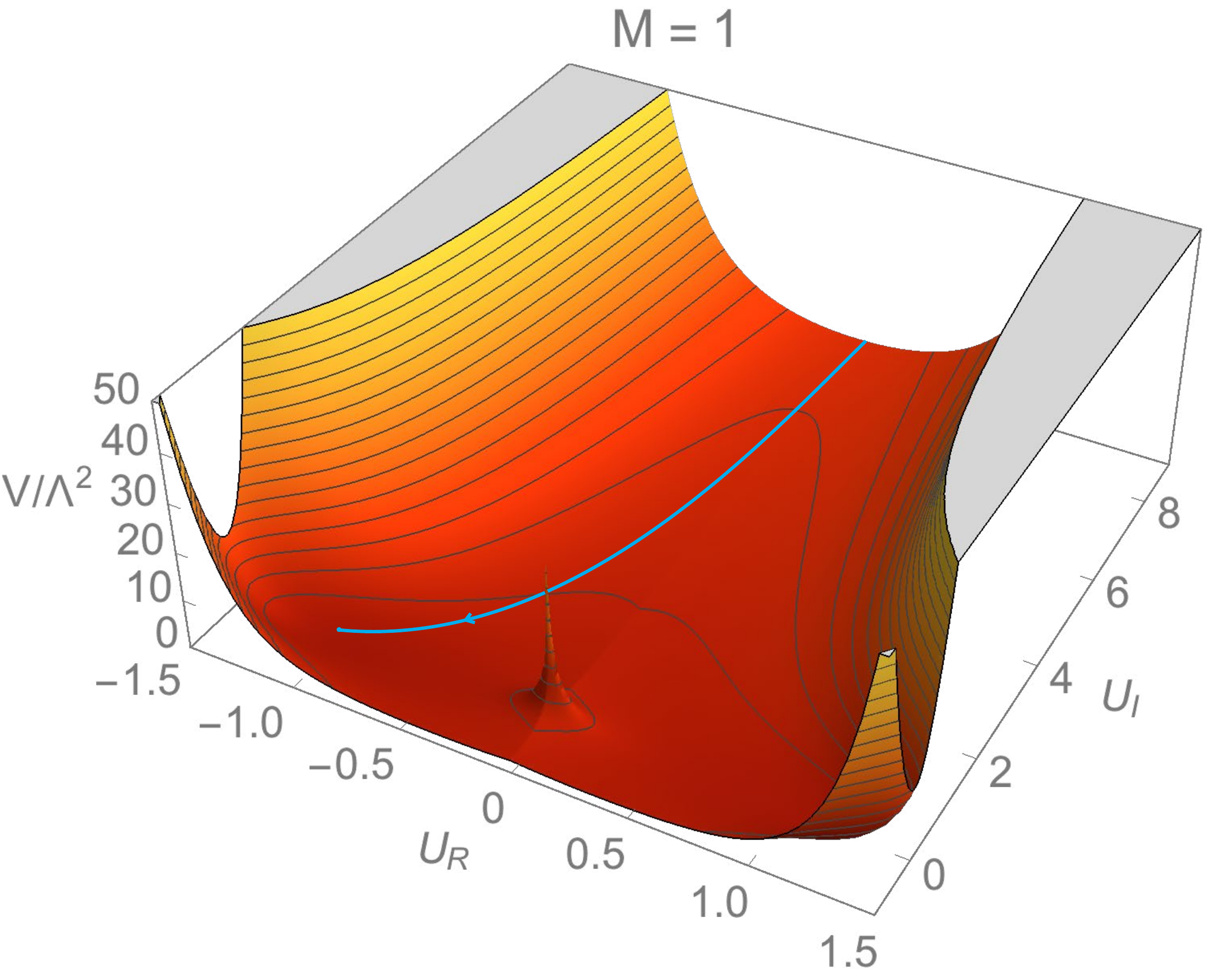}
\hspace{0.5cm}
\includegraphics[width=0.42\textwidth]{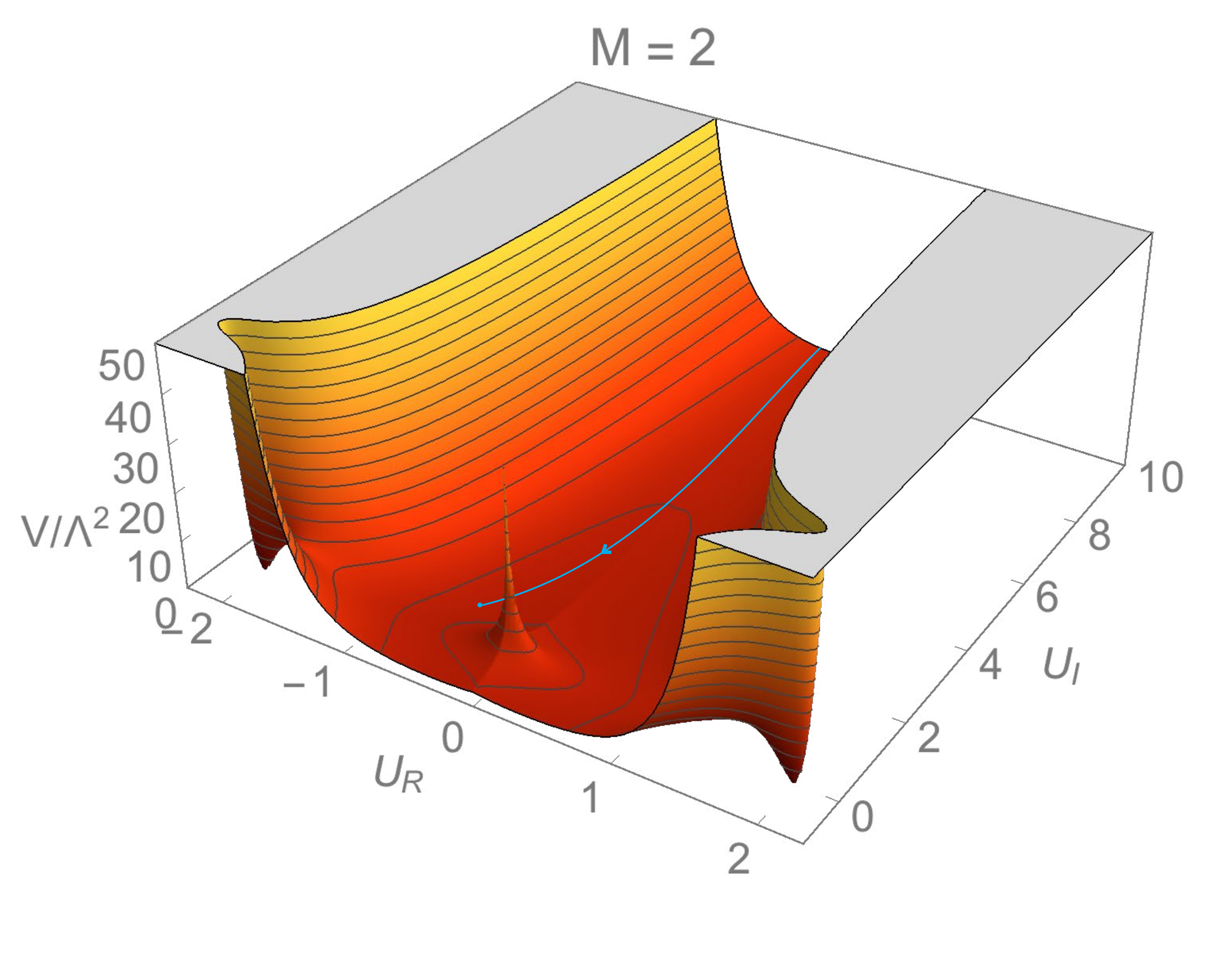}
\caption{ {Both} panels present scalar potential from  {Equation}~\eqref{eq:modelLog2} with $K = K_+$, as~a function of $u_R$ and $u_I$. During~inflation  {$|u_R| \ll M \ , \ |u_I| \gg M$} and after inflation the field rolls towards one of the minima at $u_I = 0$ and $u_R = \pm M$ (left panel). For~$M>M_c$, the~field is stuck in the  {dS} vacuum separated from the true vacuum by a local maximum (right panel).  {Blue lines present the evolution of fields $u_R$ and $u_I$ from the inflationary plateau to their minima.}}
\label{fig:V3D}
\end{figure}

This model has this interesting feature, where both $V_+$ and $V_-$ may generate successful inflation. The~scalar potential $V_+$ as a function of $u_R$ and $u_I$ is presented in the Figure~\ref{fig:V3D}. In~the case of $K = K_+$ the evolution of fields may look more complicated. First, the~field reaches the $u_R = 0$ valley, for~which the inflationary potential may be approximated as
\begin{equation}
V(u_R = 0) = \Lambda^2 \left( \log^2 \left(\frac{u_I}{M}\right) + \left(\frac{\pi}{2}\right)^2 \right) \, . \label{eq:VbigUI}
\end{equation}

\begin{figure}
\centering
\includegraphics[height=4cm]{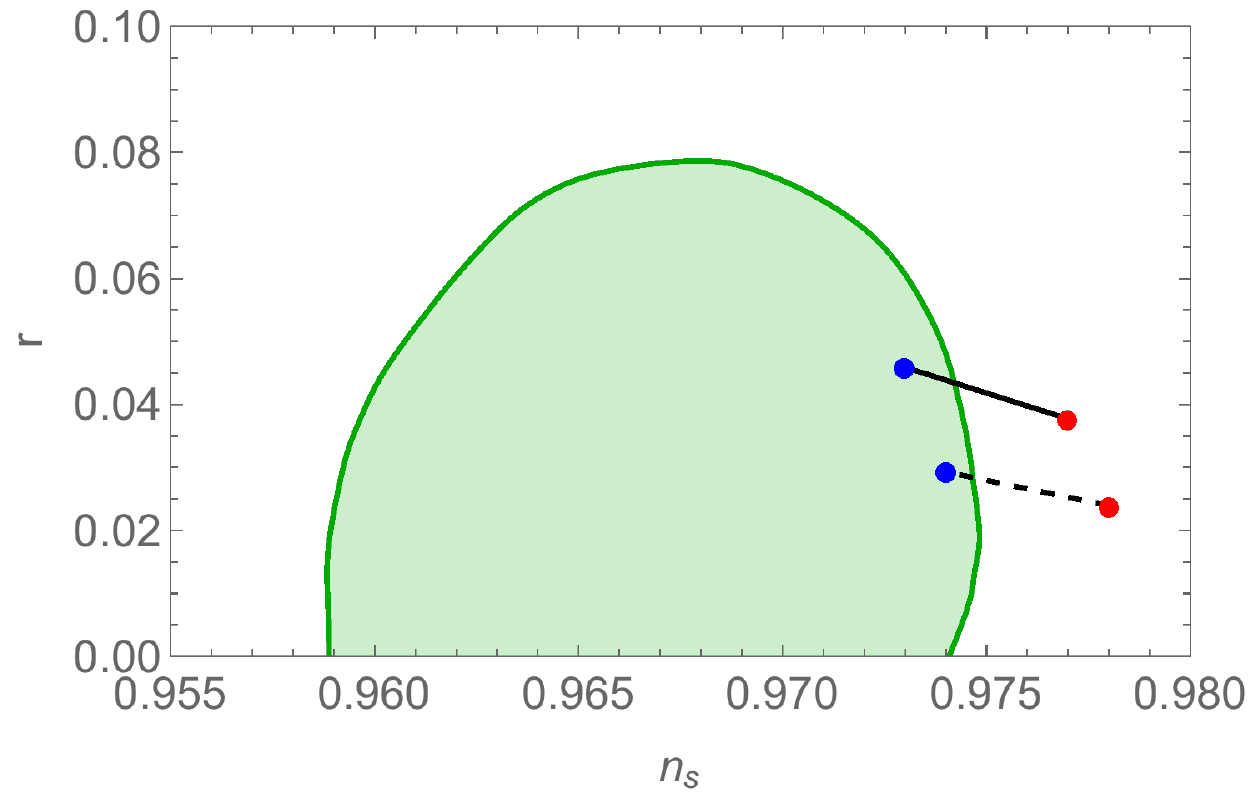}
\caption{ {Results} for the \eqref{eq:VLognKm} model for $M = 1$. Solid (dashed) lines correspond to $K=K_+$ ($K=K_-$), while red and blue dots represent $N_\star = 60$ and $N_\star = 50$ respectively. The~consistency with the data requires $N_\star \simeq 50$.
}
\label{fig:LogKm}
\end{figure}

Inflation occurs while $|u_R| \ll M$ and $u_I \gg M$. In~addition, in~order to enable the inflaton to reach the global minimum at $(u_I,u_R) = (0,\pm M)$ one requires  {$M < M_c \equiv \sqrt{e} \simeq 1.65$}. Otherwise the inflaton finds itself in a de Sitter  {(dS)} minimum at the end of the logarithmic slope and graceful exit becomes possible only via quantum tunneling. The~results of the \eqref{eq:VbigUI} model for $M = 1$ presented in the Figure~\ref{fig:LogKm} (solid line). The~potentials for both $M < M_c$ and $M > M_c$ are presented in the Figure~\ref{fig:V3D}.

\unskip

In the close vicinity of $M = M_c$ one obtains another possibility of inflation along the $u_R$ direction. To~see that let us investigate the evolution of the field for $u_I \ll M$, when inflation along the $u_I$ direction is over. In~such a case, one finds
\begin{equation}
V \propto e^{u_R^2} \log^2\left( \frac{u_R}{M} \right) \, . \label{eq:Log2saddleV}
\end{equation}

\begin{figure}
\centering
\includegraphics[width=0.45\textwidth]{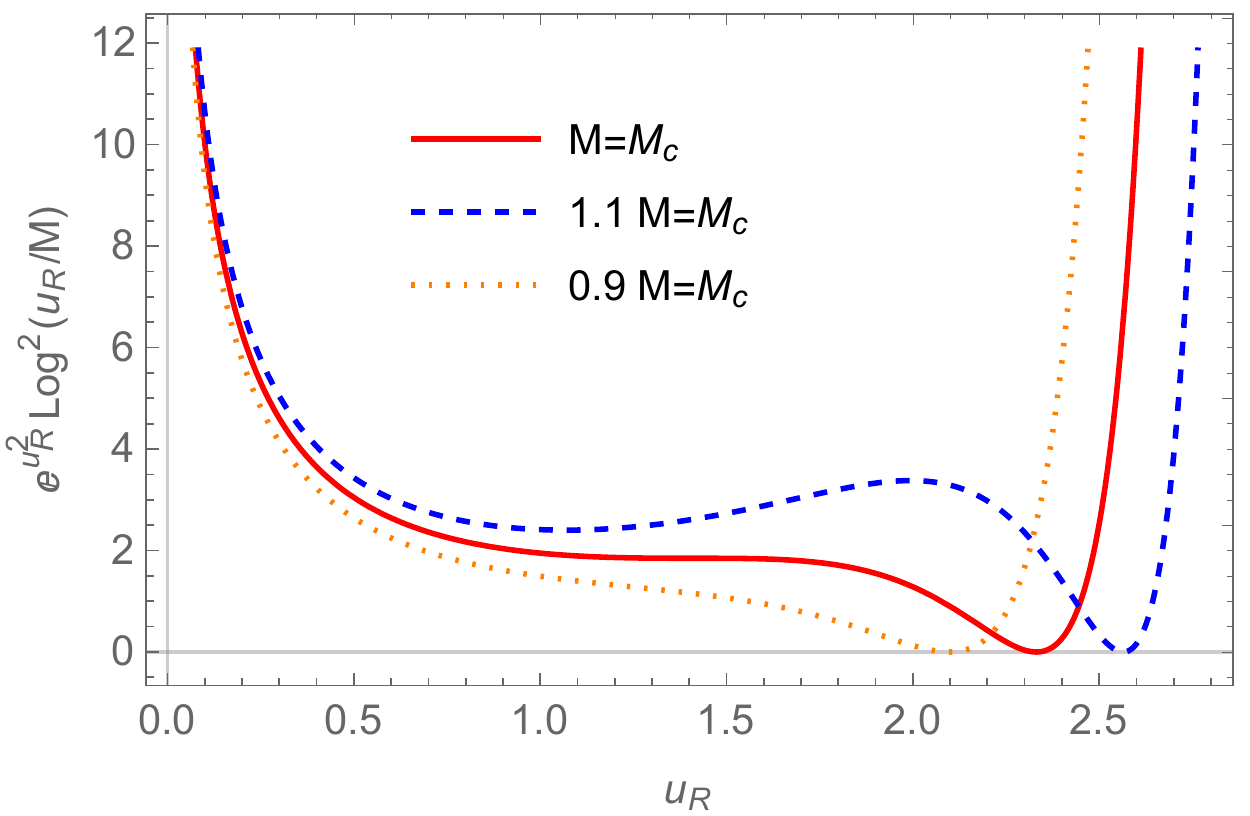}
\hspace{0.5cm}
\includegraphics[width=0.45\textwidth]{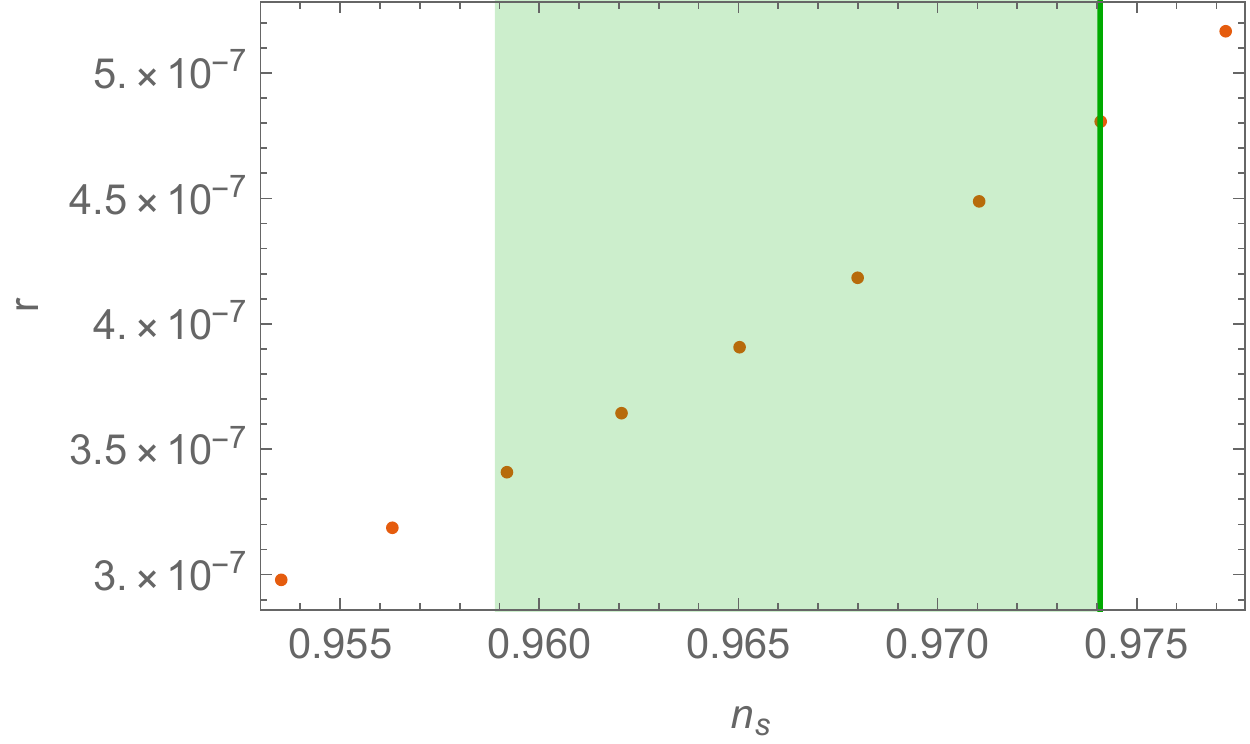}
\caption{ {Left} panel: The potential \eqref{eq:Log2saddleV} for $M \simeq M_c$. For~$M = M_c$ one obtains a saddle point, around which one obtains a second phase of inflation. For~$M > M_c$ one obtains a local maximum, which separates the inflaton from the Minkowski vacuum. Right panel: results for the \eqref{eq:Log2saddleV} on the $(n_s,r)$ plane. Dots correspond to $M = (1-k\times 5\times 10^{-6})M_c$, where $k\in \{12, 13,\ldots,20 \}$ and bigger $k$ corresponds to bigger $n_s$. Inflation can be only consistent with the data for $(1- 9.5 \times 10^{-5})M_c \lesssim M \lesssim (1- 7 \times 10^{-5})M_c$, which means that the model requires significant fine-tuning in order to be within the $2\sigma$ regime of Planck/BICEP~results.}
\label{fig:Log2Saddle}
\end{figure}

As shown in the Figure~\ref{fig:Log2Saddle}, the~potential develops a local minimum for $M > M_c$, which prevents the field to reach its global minimum. For~$M \leq M_c$ the only existing minima are in $u_R = \pm M$. For~$M = M_c$ one obtains a saddle point at $u_R = 1$, which means that inflation  {has} two phases. The~first one happens at the logarithmic slope of the $\log^2(u_I)$ inflation, while the second one happens at the plateau in the vicinity of $u_R \simeq \sqrt{2}$.

The inflation along the $u_R$ direction requires certain fine-tuning. To~obtain at least 60 e-folds one requires $M_c-M \lesssim \mathcal{O}(10^{-4})$.  
In addition, as~shown in the Figure~\ref{fig:Log2Saddle}, the~consistency with the data requires $M_c - M \simeq 10^{-5}M_c$. The~reason for the fine-tuning is the following---for the saddle-point inflation one obtains $n_s \simeq 0.92$, which is highly inconsistent with Planck/BICEP data. Thus, one must deviate from the saddle-point scenario and assume an inflection-point potential. On~the other hand, the~plateau obtained in this model is quite short, so even small deviations from the saddle-point inflation lead to insufficient number of~e-folds.

\unskip

\subsection{a Simple Plateau~Model} \label{sec:simple}

One may easily obtain a flat inflationary potential using the following field redefinition and superpotential
\begin{equation}
f = \frac{1}{\sqrt{2}}\frac{1}{T} \, , \qquad w = T - 1 \, . \label{eq:inverseplateau}
\end{equation}

In such a case the potential is equal to 
\begin{equation}
V= \Lambda ^2 e^{2 u_I^2} \left(\left(u_I^2+u_R^2\right)^{-1} - 2\left(u_I^2+u_R^2\right)^{-\frac{1}{2}} \cos \left(\arctan\left(\frac{u_I}{u_R}\right)\right)+1\right) \, .
\end{equation}

\begin{figure}
\centering
\includegraphics[width=0.45\textwidth]{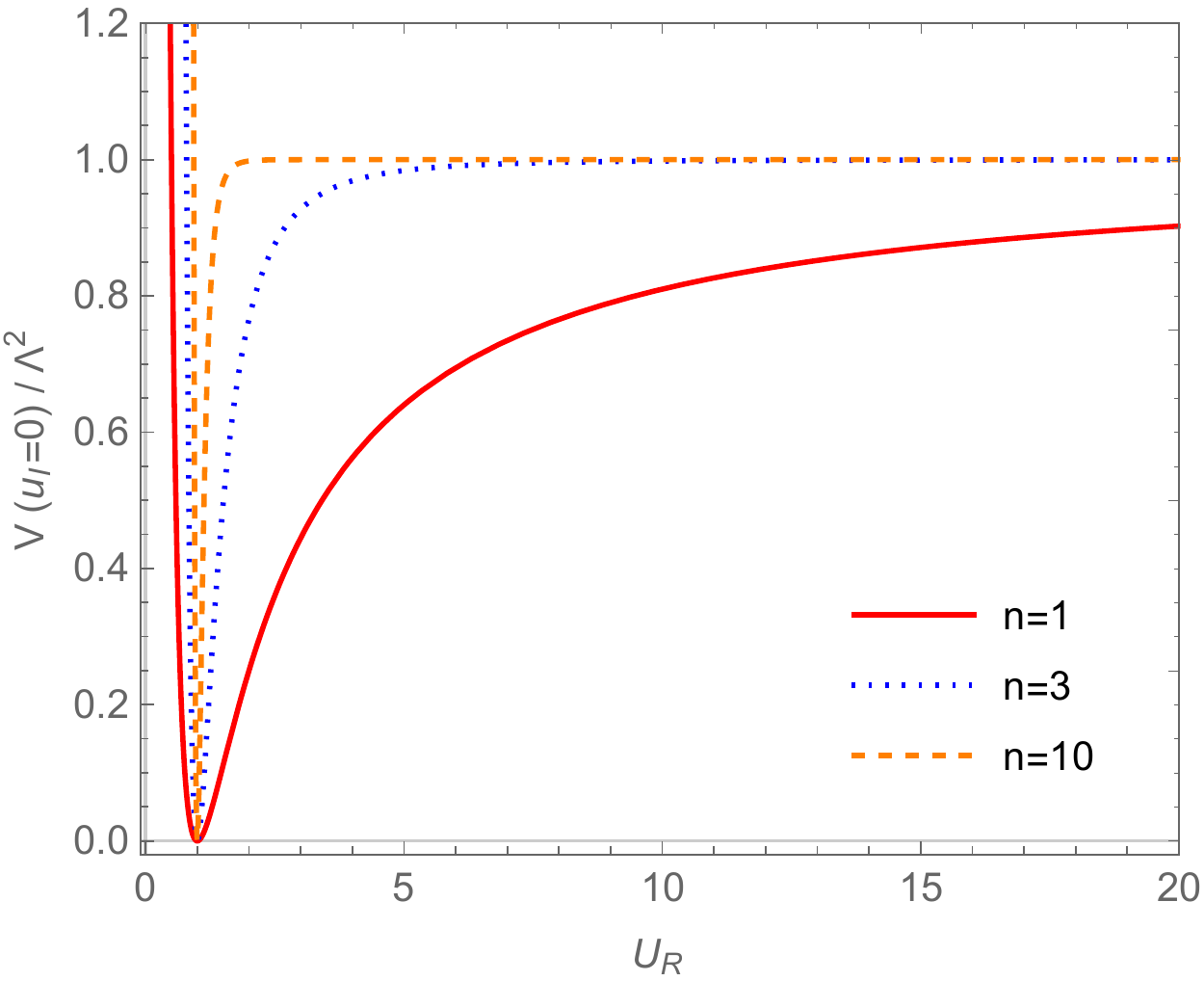}
\hspace{0.5cm}
\includegraphics[width=0.45\textwidth]{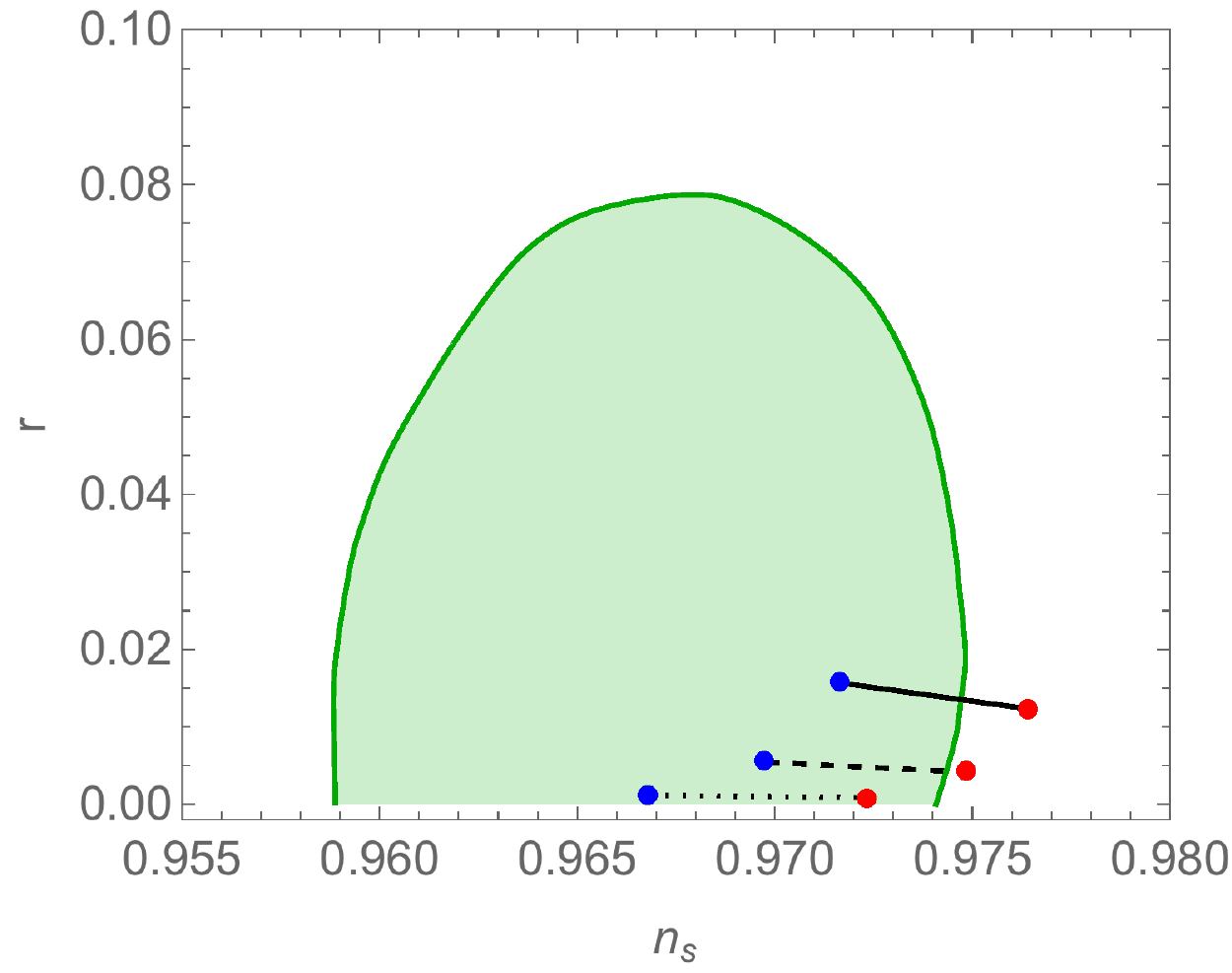}
\caption{ Left panel: The inflationary potential of model \eqref{eq:inverseplateau}  and its generalization $V = \Lambda^2 (1-u_R^{-1})^{2n}$. Right panel:  {The} tensor-to-scalar ratio and   {the} spectral index for $N_\star = 50$ and $N_\star = 60$ (solid line, blue and red dots respectively). Dashed and dotted lines represent results for the generalization of the model, for~which one has taken $f(T) = T^{-n}/\sqrt{2}$. Solid, dashed and dotted lines correspond to $n = 1$,  $n = 2$ and $n=4$ respectively.} 
\label{fig:inverseplateau1}
\end{figure}

Inflation happens in the $u_I = 0$ valley, which leads to simple inflationary potential of the form of
\begin{equation}
V(u_I = 0)  = \Lambda^2 \left( 1 - \frac{1}{u_R}\right)^2 \, . \label{eq:Vinf1inverseplateau}
\end{equation}
In such a case one finds the global minimum of the potential at $u_R = \pm 1$ and $u_I = 0$. The~scalar potential \eqref{eq:Vinf1inverseplateau} is presented in Figure~\ref{fig:inverseplateau1}. The~model can be easily generalized into $f(T) = \frac{1}{\sqrt{2}}T^{-n}$, which gives $V =\Lambda^2(1-u_R^{-1})^{2n}$. Increasing $n$ decreases $r$ and moves $n_s$ inside the $2\sigma$ of the Planck/BICEP data. The~results for $n =1$, $n = 2$ and $n=4$  {are} presented in the Figure~\ref{fig:inverseplateau1}.  {Please note that the same model could be obtained using different $f(T)$ and $w(T)$. For~instance, for~$f = e^{T}/\sqrt{2}$ and string-inspired superpotential $w = e^{-T}-1$ one obtains exactly the same form of $V(U)$.}

{ {As in the} case of Starobinsky inflation, this model could easily be embedded in a scalar-tensor theory. The~Jordan frame action
\begin{equation}
    S=\int d^4\sqrt{-g}\left(\frac{\varphi}{2}R-\frac{\omega}{\varphi}(\partial\varphi)^2-\lambda^2\varphi ^2 \left(1-\frac{\sqrt{2/\beta}}{\log \varphi}\right)^2\right)
\end{equation}
can be transformed to the Einstein frame action and potential by a standard field redefinition. This~Einstein frame potential will be fully consistent with \eqref{eq:Vinf1inverseplateau}.
}

Both spectral index and tensor-to-scalar ratio can be found analytically. Using the $N_\star \gg 1$ approximation one finds $u_{R\star} = u_R(N_\star) \simeq (2 n (n+2) N_\star)^{\frac{1}{n+2}}$, which gives

\begin{equation}
r \simeq \frac{2^5 n^2}{(2 n (n+2) N_\star)^{\frac{2 (n+1)}{n+2}}} \, ,\qquad n_s \simeq 1 -\frac{12n^2}{(2 n (n+2) N_\star)^{\frac{2 (n+1)}{n+2}}}-\frac{2 (n+1)}{(n+2) N_\star} \, .
\end{equation}

\unskip

\subsection{Plateau from the Modular~Transformation} \label{sec:modular}

Let us return to $w=T$, but~with
\begin{equation}
f = \frac{a T + b}{c T + d} \, .
\end{equation}

This form of $f(T)$ can be motivated by the modular transformation, for~which $ad - bc = 1$. For~the modular transformation one requires $a, b, c$ and $d$ to be integers. One can also consider more generic case of the $PSL(2,\mathbb{R})$ group, for~which one finds $a, b, c, d \in \mathbb{R}$. 

For $K = K_-$ and $u_I = 0$ one finds
\begin{equation}
V = \Lambda^2\left( \frac{\sqrt{2} b-d \, u_R}{\sqrt{2} a-c\,  u_R} \right)^2 \, .
\end{equation}

One can simplify this potential using $ad-bc = 1$ and a simple field transformation $u_R \to u_R +\sqrt{2} a/c$, which gives
\begin{equation}
V = \Lambda^2 \frac{d^2}{c^2} \left(1+\frac{\sqrt{2}}{c\, d}\frac{1}{u_R}\right)^2 \, . \label{eq:Vmodular1}
\end{equation}

The \eqref{eq:Vmodular1} model is simply a generalization of the \eqref{eq:Vinf1inverseplateau}. 

On the other hand, for~$K = K_+$ one finds
\begin{equation}
V(u_R = 0) = \Lambda^2 \frac{2 b^2+d^2 u_I^2}{2 a^2+c^2 u_I^2} \, . 
\end{equation}

This potential obtains a minimum for $u_I = 0$. To~obtain a Minkowski vacuum one requires $b = 0$, which together with $ad - bc = 1$ gives

\begin{equation}
V = \frac{\Lambda ^2}{a^4}\frac{\phi ^2}{ 2+ \left(\frac{c}{a}\right)^2 \phi ^2} \, . \label{eq:Vmodular2}
\end{equation}

\begin{figure}
\centering
\includegraphics[height=5cm]{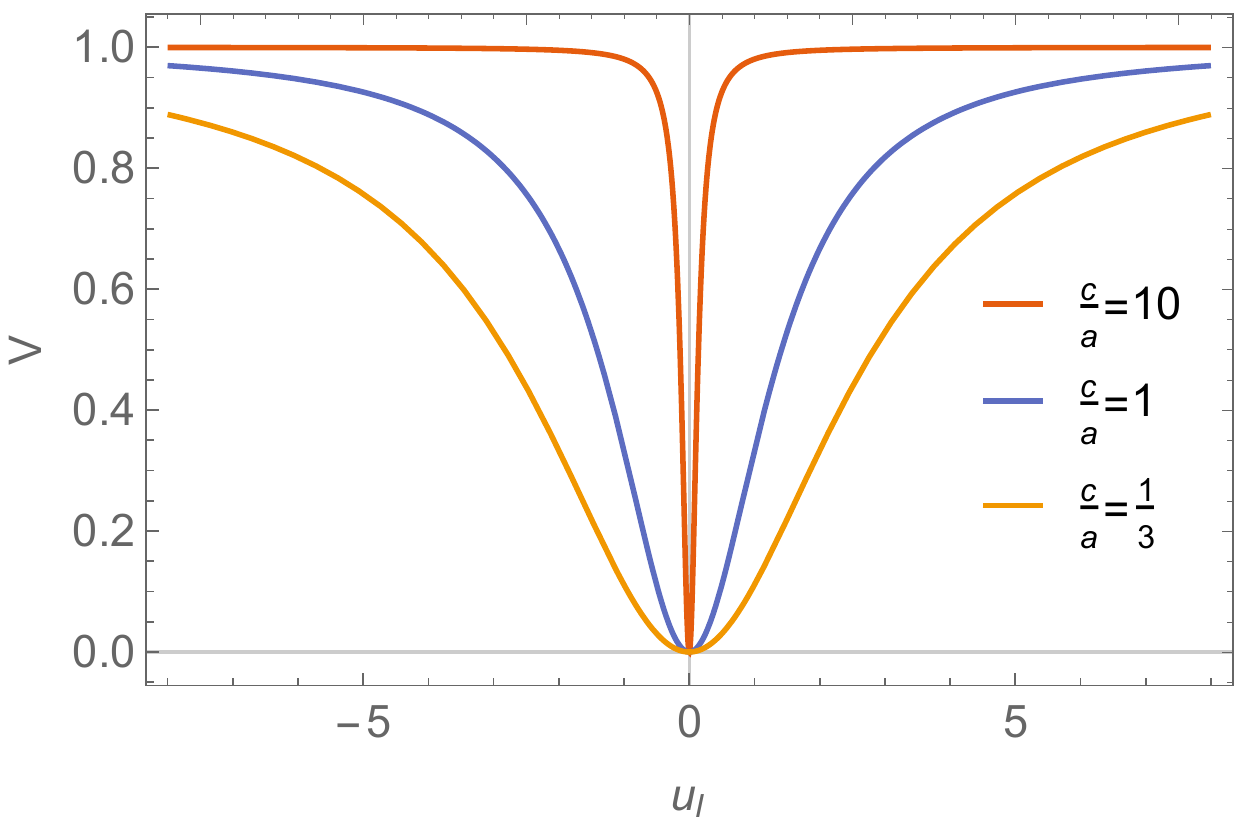}
\caption{ {Potential} \eqref{eq:Vmodular2} for different values of $c/a$. Please note that for $c \gg a$ {, inflation} takes place predominantly at the plateau, while for $c \ll a$ the last 60 e-folds of inflation happen around the $\phi^2$~slope. }
\label{fig:Vmodular2}
\end{figure}

The $a^4$ term is a part of the normalization of the potential, which could be absorbed into $\Lambda$ and does not affect $r$ and $n_s$. Thus, from~the point of view of the predictions on the $(n_s,r)$ plane, the~model has only one parameter, which is $c/a$. The~\eqref{eq:Vmodular2} potential has two limits. For~$\phi^2 \ll a^2/c^2$ one finds $V \propto \phi^2$, while for $\phi^2 \gg a^2/c^2$ the potential obtains inflationary plateau. The~potential for the \eqref{eq:Vmodular2} model is shown in the Figure~\ref{fig:Vmodular2}. Assuming the slow-roll approximation one finds
\begin{eqnarray}
r &\simeq& \frac{16}{\left(\frac{a^2}{c^2}+4 N_\star\right) \left(\sqrt{\frac{4 c^2 N_\star}{a^2}+1}-1\right)} \, ,\nonumber \\
 n_s &\simeq& 1 -\frac{1}{N_\star}\left(\frac{1}{\sqrt{\frac{4 c^2 N_\star}{a^2}+1}}+1\right)-\frac{2}{\frac{a^2}{c^2}+4 N_\star} \, .
\end{eqnarray}

In the small/big $4N_\star c^2/a^2$ limit $r$ and $n_s$ can be  further simplified to
\begin{eqnarray}
r \simeq \left|\frac{a}{c}\right| \frac{2}{N_\star^{3/2}} \, , \qquad n_s \simeq 1 - \frac{3}{2N_\star} \qquad &\text{for}& \qquad 4N_\star c^2/a^2\gg 1\, ,\\
r \simeq \frac{8}{N_\star} \, , \qquad n_s \simeq 1 - \frac{2}{N_\star} \qquad &\text{for}& \qquad 4N_\star c^2/a^2\ll 1 \, .
\end{eqnarray}

\begin{figure}
\centering
\includegraphics[width=0.45\textwidth]{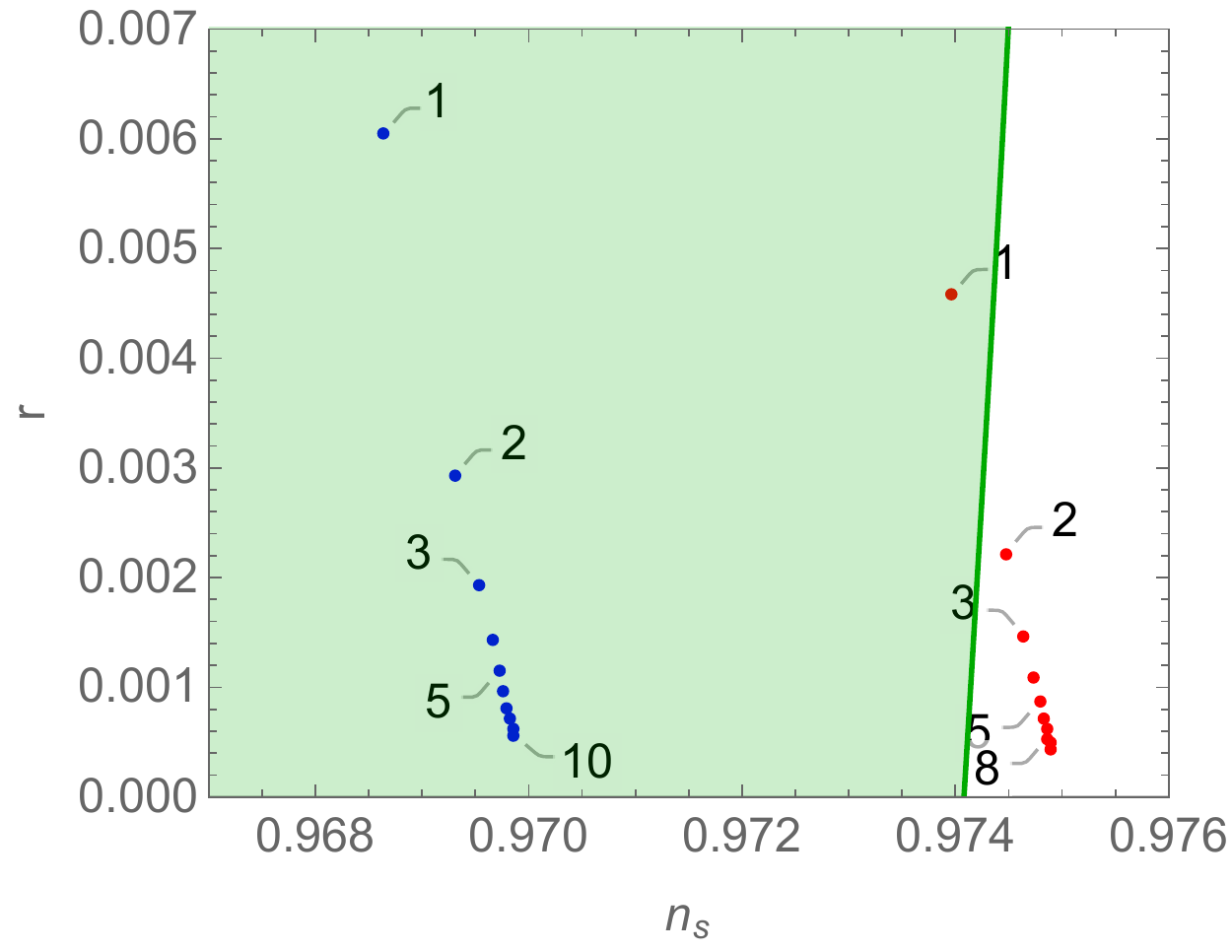}
\hspace{0.5cm}
\includegraphics[width=0.45\textwidth]{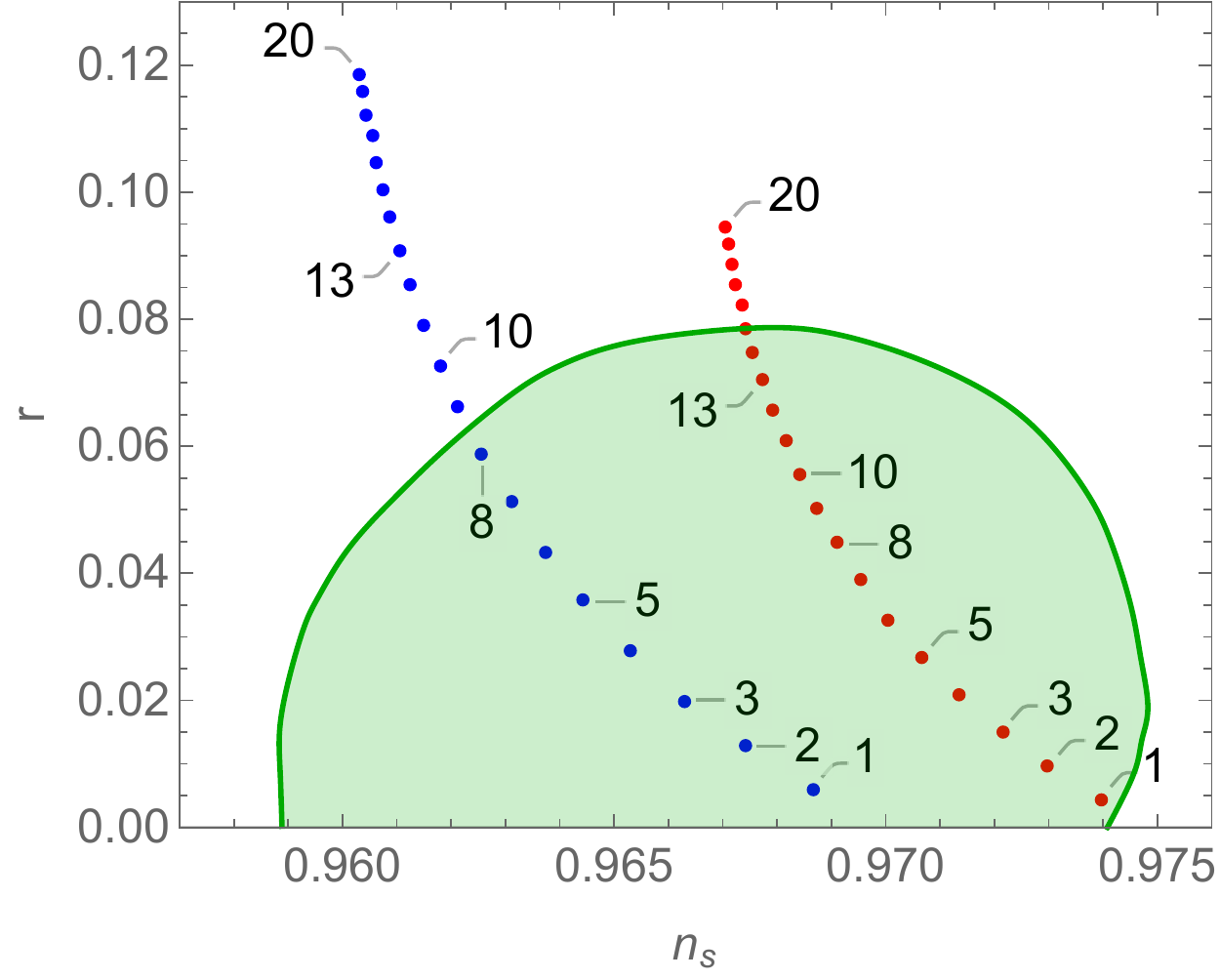}
\caption{ {Results} for the \eqref{eq:Vmodular2} model for both $c/a>1$ and $c/a < 1$. Blue and red dots represent $N = 50$ and $N = 60$ respectively. Left panel: We consider $k = c/a$ and $k\in\{1,2,\ldots,10\}$. Right panel: $k = a/c$ and $k\in\{1,2,\ldots,20\}$. Note how in the $c\ll a$ limit one moves towards the results of $\phi^2$ inflation.}
\label{fig:rnsmodular2}
\end{figure}

In the $4N_\star c^2/a^2\ll 1$ one finds $r \sim 10^{-1}$, which is inconsistent with PLANCK/Bicep data. The~comparison to the Planck/BICEP data is presented in Figure~\ref{fig:rnsmodular2}.

\unskip

\unskip

\subsection{Bell-Curve~Potentials} \label{sec:bell}

Let us consider
\begin{equation}
f = -\frac{1}{\sqrt{2}}M \log^p(T) \, , \qquad w = T \, ,
\end{equation}
which gives the scalar potential of the form of
\begin{equation}
V = \Lambda^2 \exp \left( u_I^2 -2 \left( \frac{u_R^2 + u_I^2}{M^2} \right)^{\frac{1}{2p}}\cos \left( \frac{1}{p}\arctan \frac{u_I}{u_R} \right)  \right) \, . \label{eq:Vanyn}
\end{equation}

{Inflation happens for $u_I = 0$, which for a generic value of $p$ gives the following inflationary potential}
\begin{equation}
V(u_I = 0) = \Lambda^2 \exp \left(-\left| \frac{u_R}{M} \right|^{1/p} \right) \, . \label{eq:Vexpinfl}
\end{equation}

\begin{figure}
\centering
\includegraphics[width=0.45\textwidth]{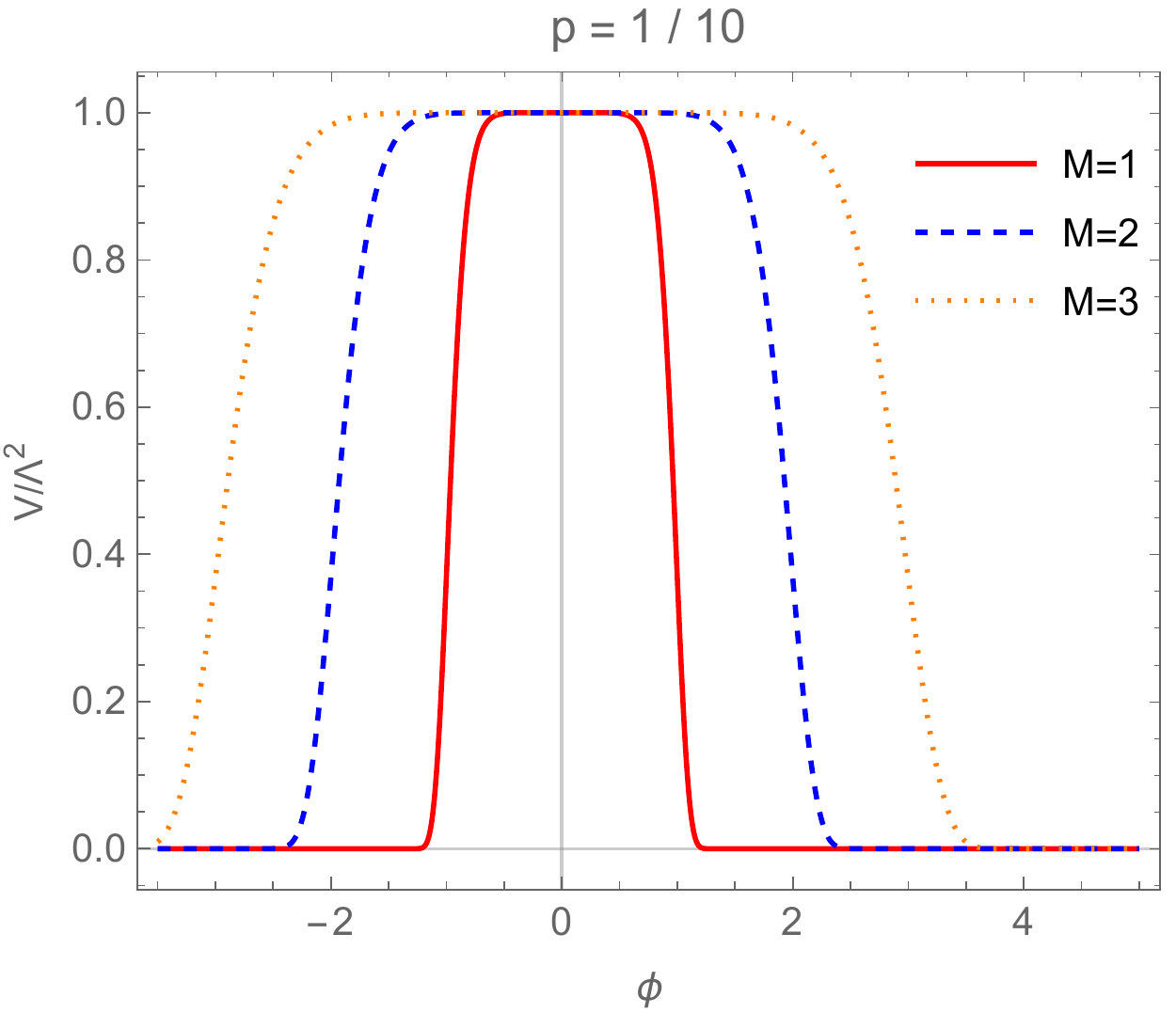}
\hspace{0.5cm}
\includegraphics[width=0.45\textwidth]{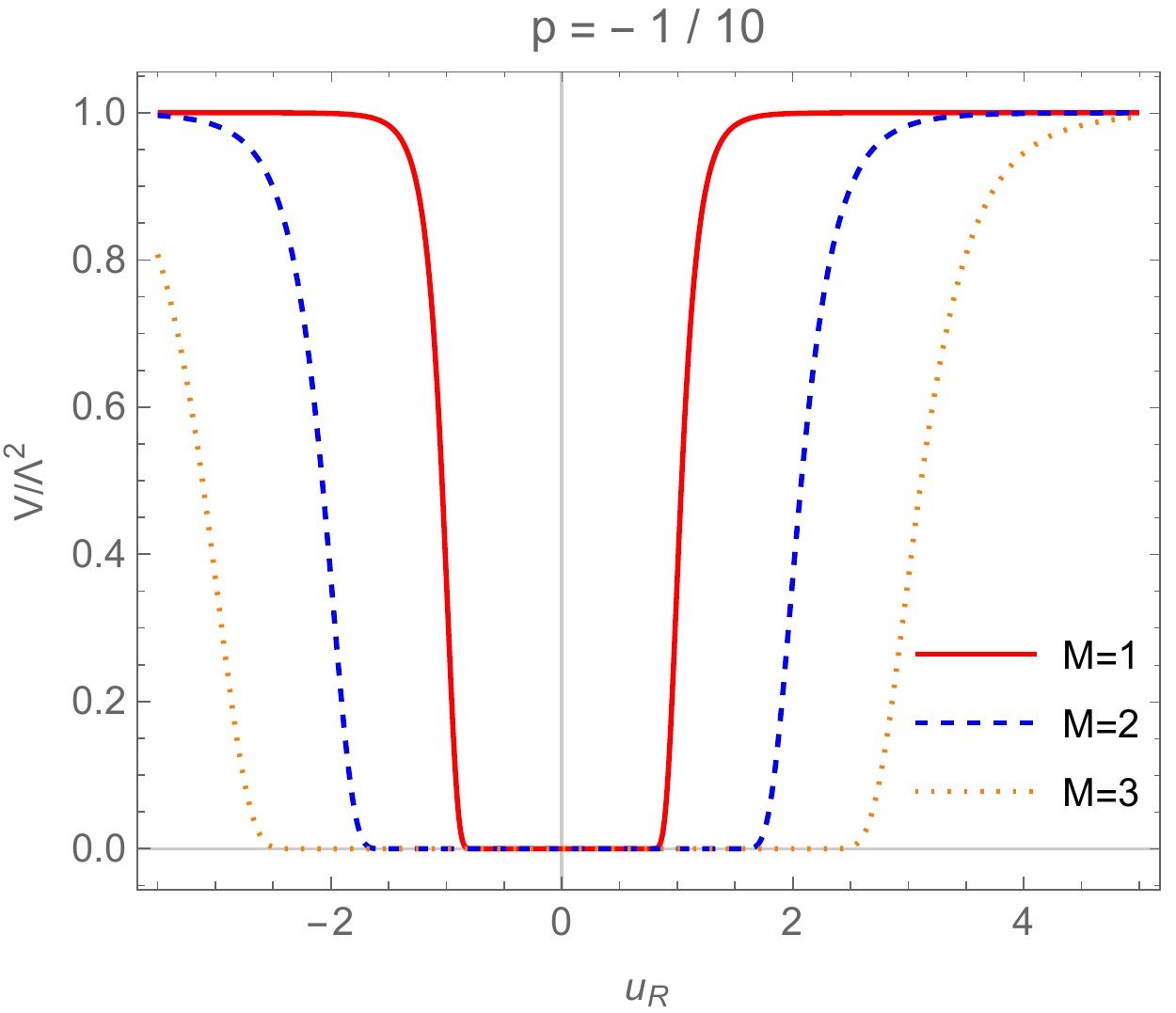}
\caption{ {Potential} \eqref{eq:Vexpinfl} for $p > 0$ and $p < 0$ (left and right panel respectively).  For~$p > 0$ inflation happens around the local maximum and the potential does not have a minimum. For~$p < 0$ inflation happens on one of the infinite plateaus separated by the~minimum.}
\label{fig:Vanynlimits}
\end{figure}

The shape of the potential for $p>0$ and $p<0$ is presented in the Figure~\ref{fig:Vanynlimits}.
  {This model could also be expressed using a string-inspired superpotential $w = e^{-T}$ together with $f = M T^p/\sqrt{2}$. }
 Within the slow-roll approximation one finds
\begin{eqnarray}
r &\simeq& 16 \left(\frac{2 N_\star (1-2 p)}{\left(\sqrt{2} M |p|\right)^{\frac{1}{1-p}}}+1\right)^{-\frac{2 (1-p)}{1-2 p}} \, \label{eq:rstar}\\
n_s &\simeq& 1 - 2 \left(\frac{2 N_\star (1-2 p)}{\left(\sqrt{2} M |p|\right)^{\frac{1}{1-p}}}+1\right)^{-\frac{2 (1-p)}{1-2 p}}-\frac{4 (1-p)}{\left(\sqrt{2} M |p|\right)^{\frac{1}{1-p}}+2 N_\star (1-2 p)} \, . \label{eq:nsstar}
\end{eqnarray}

Please note that in the
\begin{equation}
N_\star \ll \frac{1}{2 (1-2 p)} \left(\sqrt{2} M |p|\right)^{\frac{1}{1-p}}
\end{equation}
regime  {equation} \eqref{eq:rstar} takes the form of
\begin{equation}
\epsilon \simeq \frac{\epsilon_0}{N^n} \, ,\label{eq:epsilonparametrisation}
\end{equation}
where $M = (2 p-1) \sqrt{\epsilon_0} (2 (1-2 p) \epsilon_0)^{-p}p^{-1}$ and $n=2(p-1)/(2p-1)$. $0<p<1/2$ corresponds to $n > 2$, while $p<0$ to $1<n<2$. Please note that the $p \to 0$ limit corresponds to $\epsilon \propto N^{-2}$, which is the feature of theories like Starobinsky inflation, Higgs inflation, or~$\alpha$-attractors. Indeed, for~$|p|M \ll 1$ one finds
\begin{equation}
r \simeq 8\frac{M^2p^2}{N_\star^2} \, , \qquad n_s \simeq 1-\frac{2}{N_\star} \, ,
\end{equation}
which is the result of the \eqref{eq:VStarobinskyapp} model for $M \to M |p|$.

\begin{figure}
\centering
\includegraphics[height=5cm]{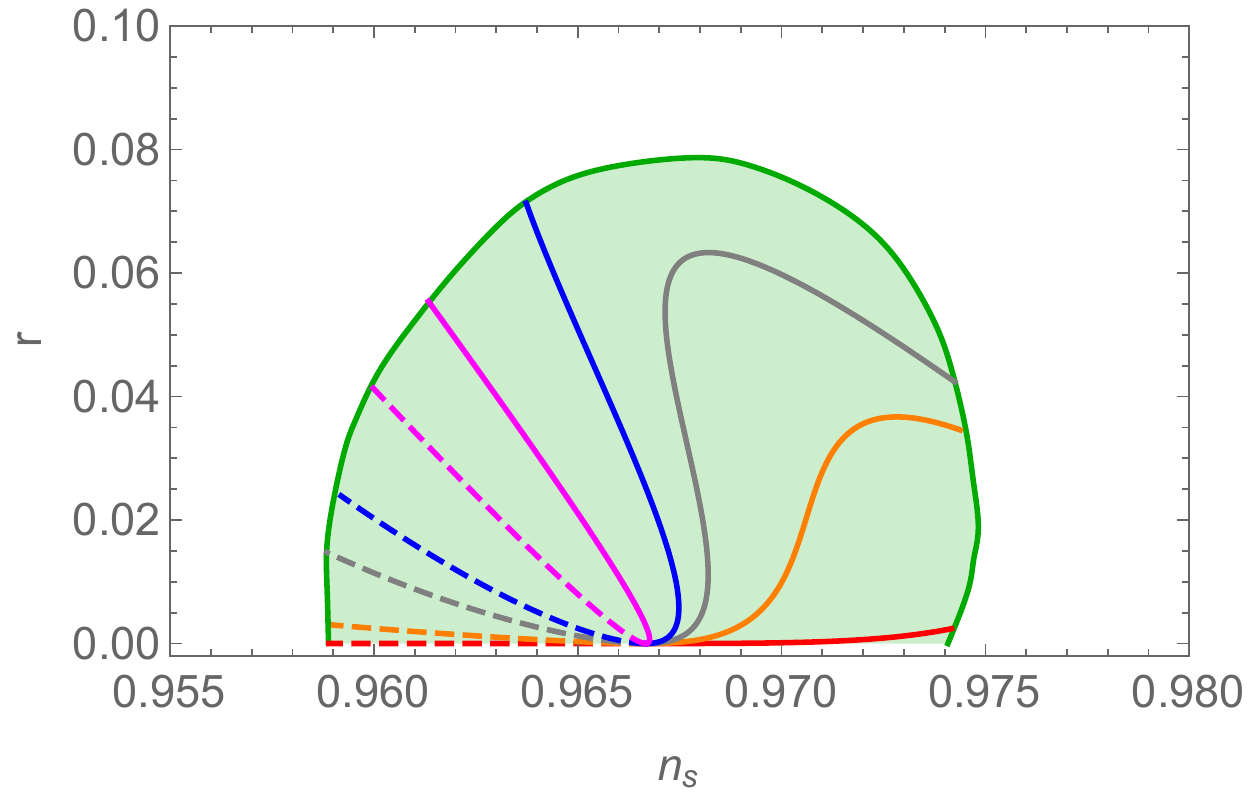}
\caption{ {Results} for the \eqref{eq:Vexpinfl} model for $N_\star = 60$ and $p < 0$ or $p > 0$ (solid and dashed lines respectively). Red, orange, gray, blue and pink lines represent $M = 1$, $M = 10$, $M = 20$, \mbox{$M = 30$} and $M = 100$ and $p \in (-0.44,0.16)$, $p \in (-1.3,0.152)$, $p \in (-1.7,0.132)$,  $p \in (-0.28,0.106)$ and \mbox{$p \in (-0.053,0.042)$} respectively. Note how results for different $M$ reproduce the result of the Starobinsky inflation in the $p \to 0$ limit. In~addition, the~results of the \eqref{eq:Vexpinfl} model may cover the whole $2\sigma$ regime of the Planck/BICEP~data.}
\label{fig:VBellrns}
\end{figure}

The \eqref{eq:epsilonparametrisation} parametrization is widely used in cosmology (see e.g.,~\cite{Mukhanov:2014uwa}), which is an additional motivation to consider this model. We have assumed $n \neq 1$ and $n \neq 2$, since $n=1$ or $n=2$ correspond to well-known cases of the power-law and Starobinsky potentials, respectively. One can strongly constrain the possible range of $n$ (and consequently, the~range of $p$) using the Planck/BICEP data. Since $\frac{dN}{d\phi} = 1/\sqrt{2\epsilon}$, one finds
\begin{equation}
\frac{d\sqrt{2\epsilon}}{d\phi} = \pm\frac{1}{\sqrt{2\epsilon}}\frac{d\sqrt{2\epsilon}}{dN} = \pm \frac{\epsilon_N}{2\epsilon} \, ,
\end{equation}
as well as
\begin{equation}
\frac{d\sqrt{2\epsilon}}{d\phi} = \pm \frac{d}{d\phi}\frac{V_\phi}{V} = \pm \left( \eta - 2\epsilon \right)\, ,
\end{equation}
where $\epsilon_N = \frac{d\epsilon}{dN}$. Therefore, one can express $\eta$ as
\begin{equation}
\eta = 2\epsilon + \frac{\epsilon_N}{2\epsilon} \, ,\qquad n_s = 1-6\epsilon + 2\eta = 1 + \frac{\epsilon_N}{2\epsilon} - 2\epsilon \, .
\end{equation}
Using the \eqref{eq:epsilonparametrisation} one finds
\begin{equation}
\eta = 2\frac{\epsilon_0}{N^n} \, , \qquad n_s = 1 - \frac{n}{N} - 2\frac{\epsilon_0}{N^n}\, .
\end{equation}
Since we require $n_s$ to be in the vicinity of $0.96$ one cannot obtain a result consistent with Planck/BICEP data for $n$ significantly bigger than 2. The~results for the \eqref{eq:Vexpinfl} model  {are presented in } Figure~\ref{fig:VBellrns}.

\unskip

 {Let us discuss possible reheating mechanisms in this scenario. In general, for any $p$ reheating may happen at the post-inflationary slope, e.g.,~via instant preheating}~\cite{Felder:1998vq}  {and for~$p<0$ through the standard reheating oscillations. However, in our case we have a built-in gravitational particle production}~\cite{Ford:1986sy,Artymowski:2017pua}  {for any $p$ if our inflaton is of the dark sector. This is due to the steepness of the post-inflationary potential. In~the gravitational reheating case the post-inflationary evolution is dominated by the inflaton with an equation of state $w=1$. Reheating (understood as the beginning of the radiation domination) happens around $T_R \sim \mathcal{O}(10^7GeV)$. Hence, this is another prediction of the model, which narrows down possible values of $N_\star$ and $n_s$.} 

To conclude, the~\eqref{eq:Vexpinfl} model has several interesting features, which are worth exploring in detail. The~model spans the entire allowed $(n_s,r)$ region, which makes its predictions consistent with any future experiment that would constrain these parameters. { {This means that inflation can have an arbitrarily low scale, which is strongly favored by the  trans-Planckian censorship conjecture}~\cite{Brahma:2019vpl,Cai:2019dzj,Berera:2020dvn}}. 
 {The model} also contains an inflationary attractor in the $p \to 0$ limit, which is Starobinsky-like. In~addition, the~model has a  {built-in predictive} mechanism of gravitational reheating, due to the post-inflationary kination of the inflaton. Thus, one does not require any additional assumptions regarding the couplings of the inflaton to matter~fields. 

\section{Conclusions} \label{sec:concl}

The main idea of this paper is to show that assuming a particular flat K\"ahler potential $K_{\pm} = \pm\frac{1}{2}(T\pm \bar{T})^2 + S\bar{S}$ and a specific superpotential $W = \Lambda S \, T$ one can obtain a wide spectrum of inflationary theories differentiated only by a field redefinition applied solely to the K\"ahler potential $K(f(T),\bar{f} (\bar{T}),S\bar{S})$ as introduced in Section~\ref{sec:general}. A~canonical K\"ahler potential may be restored after the field redefinition $U = f(T)$. Following this transformation of the K\"ahler potential, a~generic form of a scalar potential has been obtained, while the K\"ahler geometry remained~flat. 

In a few simple examples, we have demonstrated completely different classes of models, small~field, Starobinsky, large field and even a squared logarithmic potential as suggested in~\cite{Ben-Dayan:2018mhe}. {In~Sections~\ref{sec:monomial}--\ref{sec:starobinsky} we have presented the results of well-known inflationary models. Our goal is to emphasize that one can obtain them in a novel way using a flat K\"ahler geometry and  {a simple fixed} superpotential $W=\Lambda \, S \, T$.} We have also constructed the plateau model from a modular or PSL(2,R) transformation that covers most of the allowed parameter space. The~predictions of investigated  {models} span $r\in[10^{-6},0.06]$, and~a valid $n_s$ for reasonable number of e-folds.  Of~particular interest are the bell-curve models since their predictions cover all allowed values of $r$ and $n_s$. In~addition, due to the kination of the inflaton after inflation, the~model has a  {built-in predictive} mechanism of gravitational reheating, which does not require additional couplings of $T$ to matter fields. The~specific details of  {the analyzed models are given in} Section~\ref{sec:prototypes}.

The idea of the field redefinition bypasses the limit of holomorphicity used for generating general potentials by having an arbitrary holomorphic function $f(T)$ in $W$. This is because at the fundamental level, with~the $T$ field, we have the same holomorphic $W$, and~what one really has are different real Kahler potentials, but~all with the same Kahler geometry. As~such, the~construction abides the standard rules of SUGRA models. 
We have focused on the F-term scalar potential in SUGRA. It would be interesting to apply this method for D-term inflationary model building as~well.

\end{document}